\documentclass[journal]{IEEEtran}

\IEEEoverridecommandlockouts
\usepackage{cite}
\usepackage{graphicx}
\usepackage{textcomp}
\usepackage{xcolor}
\usepackage{outlines}
\usepackage{dsfont}

\def\BibTeX{{\rm B\kern-.05em{\sc i\kern-.025em b}\kern-.08em
		T\kern-.1667em\lower.7ex\hbox{E}\kern-.125emX}}

\usepackage{standalone}
\usepackage[reqno]{amsmath}
\usepackage{amssymb}

\usepackage{algorithm}
\usepackage{algorithmicx}
\usepackage{algpseudocode}
\usepackage{setspace}
\usepackage{multicol}
\usepackage{enumerate}
\usepackage[font=small, belowskip=-15pt,aboveskip=0pt]{caption}
\usepackage{cite}
\usepackage{amsthm}
\usepackage{cleveref}

\usepackage[]{graphicx}
\usepackage{tikz}
\usepackage{tkz-tab}
\usepackage{pgfplots}
\usepackage{booktabs}
\usepackage{floatflt}

\usepackage{psfrag}	
\usepackage{array}
\usepackage{multirow,hhline}
\usepackage{exscale}
\usepackage{color}
\usepackage{colortbl}
\usepackage{epsfig}
\usepackage{epstopdf}

\usepackage{array}
\usepackage{subcaption}

\usepackage[latin1]{inputenc} 
\usepackage[T1]{fontenc} 
\usepackage{rotating}
\usepackage{pbox}

\usepackage{pifont}

\usepackage{comment}
\usepackage[normalem]{ulem}
\usepackage{cancel}

\usepackage{bigints}

\usetikzlibrary{shapes,snakes}
\usetikzlibrary{calc}
\usetikzlibrary{patterns}
\usetikzlibrary{decorations.pathmorphing} 
\usetikzlibrary{spy}
\usetikzlibrary{positioning}
\usetikzlibrary{arrows, decorations.markings}
\usetikzlibrary{shapes.multipart, shapes.geometric, fit, backgrounds}
\usetikzlibrary{spy}

\definecolor{mycolor1}{rgb}{0.00000,0.44700,0.74100}%
\definecolor{mycolor2}{rgb}{0.85000,0.32500,0.09800}%
\definecolor{mycolor4}{rgb}{0.92900,0.69400,0.12500}%
\definecolor{mycolor3}{rgb}{0.49400,0.18400,0.55600}%
\definecolor{mycolor5}{rgb}{0.46600,0.67400,0.18800}%
\definecolor{mycolor6}{rgb}{0.30100,0.74500,0.93300}%
\definecolor{mycolor7}{rgb}{0.63500,0.07800,0.18400}%

\title{Resilience and Criticality: Brothers in Arms for 6G}
\begin{document}
 
	\author{
		\IEEEauthorblockN{Robert-Jeron Reifert,
			Yasemin Karacora,
			Christina Chaccour,
			Aydin Sezgin, and
			Walid Saad
	}
    \thanks{This work has been submitted to the IEEE for possible publication. Copyright may be transferred without notice, after which this version may no longer be accessible. \newline
	\IEEEauthorblockA{Robert-Jeron Reifert, Yasemin Karacora, and Aydin Sezgin are with Digital Communication Systems, Ruhr University Bochum, Bochum, Germany. (Email: \{robert-.reifert,yasemin.karacora,aydin.sezgin\}@rub.de).\newline
	Christina Chaccour is with Ericsson Inc., Plano, Texas, USA. (Email: christina.chaccour@ericsson.com).\newline
	Walid Saad is with the Bradley Department of Electrical and Computer Engineering, Virginia Tech, Arlington, Virginia, USA. (Email: walids@vt.edu).}}
    }

	\markboth{Preprint Submitted to the IEEE -- Resilience and Criticality: Brothers in Arms for 6G}%
	{}
	
	\maketitle
 
\begin{abstract}
    In this paper, we develop the first comprehensive tutorial on designing future 6G networks that synergistically integrate notions of resilience and criticality from the ground up.
	While resilience refers to the ability to absorb, adapt to, and recover from adversarial or challenging conditions, criticality indicates the degree of importance or urgency assigned to a particular service or component.
	Despite a spiking interest in designing resilient wireless networks, most prior works do not provide a unified resilience definition, nor harness the intricate interplay between resilience and criticality. In order to fill this gap, in this paper, we highlight the importance of a criticality-aware approach as a key enabler for providing reliable and resilient service functionality. 
	Moreover, we delve into the unique challenges and opportunities of the envisioned 6G features pertaining to resilience and (mixed) criticality. 
	After reviewing resilience definitions, we present a core resilience strategy, a unified resilience metric, different criteria for service criticality, and prioritization frameworks, that augment the 6G resilience prospects. 
    Afterwards, we explore the opportunities presented by promising technologies that enable a resilient 6G network design from a radio access network protocol stack perspective. We briefly revisit state-of-the-art network architectures, establish a rough connection to the Open-RAN Alliance vision, and discuss opportunities, existing techniques, and promising enabling mechanisms for 6G at each layer.
	Finally, the article discusses important research directions and open problems concerning resilience and criticality in 6G.
 \end{abstract}
	
	\section{Introduction}
	\label{secrel}
	Recent advances and global research activities have set the stage for future 6G networks \cite{you2021towards}. 6G aims to provide a fully fledged user experience with applications that are either human-centric or artificial intelligence (AI)-based, supporting connected intelligence. Designing 6G networks becomes increasingly challenging, as emerging 6G applications not only require extremely high quality of service (QoS), but also stringent compliance with those demands even in face of disturbances, resulting from various sources such as failed network components, channel outages, or malicious attacks. Further, many anticipated 6G  applications impose increasingly rigorous reliability demands and are highly time critical; prime examples include sectors such as manufacturing, surgical procedures, and critical infrastructures, 
    where communication interruptions may lead to hazardous situations endangering lives and causing environmental harm. Given that high-data-rate services like extended reality (XR) are likely to be deployed in these contexts, the gravity of service outages and performance degradation becomes even more pronounced \cite{chaccour2020ruin}.
	However, remarkably, to date, ongoing 5G and 6G research have focused primarily on the measures of reliability and robustness \cite{7894280,s22030762}. Specifically, a reliable network provides the required service without malfunctioning (with minimal disruptions) over a specific period, while robustness refers to withstanding disturbances and maintaining an acceptable performance level. Through the implementation of 5G, it became evident that resilience could often be ensured by building upon robustness and reliability through stringent requirements tailored to specific use cases \cite{electronics11030412}. This approach was sufficient given the relative simplicity of 5G use cases, where differentiated connectivity and QoS parameters provided adequate solutions. However, as we move toward 6G, the landscape is evolving dramatically. The increasing complexity of use cases--spanning differentiated quality of experience (QoE) needs, diverse application requirements, and the emergence of AI-native systems--demands a shift from ad hoc resilience measures to \emph{resilience-by-design}.\looseness-1

    In 6G, the network's role is no longer limited to enabling reliable connectivity; it becomes a critical enabler of intelligence flow between systems. This transformative role introduces unprecedented challenges, as the network must not just support connectivity but also adaptive, context-aware, and autonomous behaviors. The unpredictable nature of potential disruptions and the intricate interdependencies of AI-native applications further exacerbate these challenges, making traditional reliability and robustness measures insufficient.
    \emph{Resilience} in 6G must go beyond absorbing disturbances and ensuring operational stability. It requires a comprehensive approach that integrates adaptive mechanisms to reorganize and maintain functionality under diverse and evolving conditions \cite{rak2020,STERBENZ20101245,Cassottana2023}. As defined by the International Telecommunication Union (ITU), resilience is now a key design, deployment, and operational consideration for 6G systems, enabling them to continue operating and recover swiftly from disruptive events \cite{IMT-2030}. This fundamental shift underscores the necessity of embedding resilience into the fabric of 6G networks to address the complexity, intelligence, and dynamic requirements of next-generation applications.
    
\begin{figure}[!t]
	   \centering
	   \includegraphics[width=.9\linewidth]{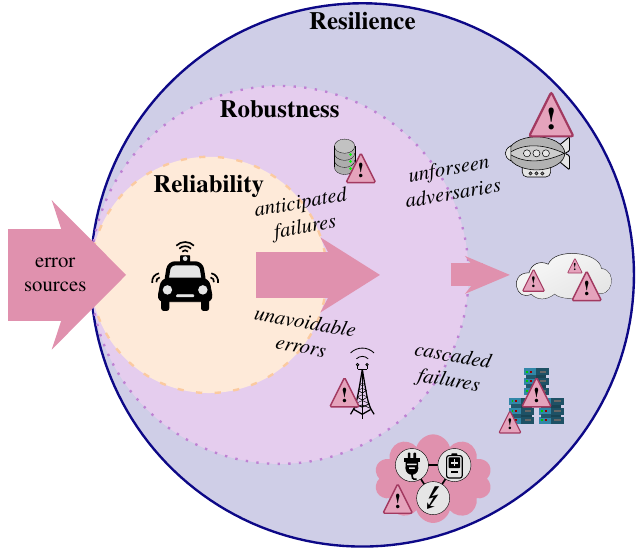}
	   \caption{Conceptual illustration of reliability (failure prevention), robustness (handling anticipated errors), and resilience (managing unforeseen adversities and cascading failures).}
	   \label{fig:fig_1_circles_v4}
    \end{figure}

    In essence, resilience enables a system to autonomously adapt to erroneous influences, recover in a timely manner, and thereby guarantee proper operation in the face of errors. Resilience goes beyond reliability and robustness in a sense that reliability aims to prevent failures, robustness focuses on performing well during anticipated disturbances, and resilience emphasizes the network's ability to seamlessly cope with and recover from unforseen failures and disruptions, as illustrated in Fig.~\ref{fig:fig_1_circles_v4}. \emph{Reliability} is a long-term metric assessing the absence of failures; \emph{robustness} prepares systems for known uncertainties; and \emph{resilience} addresses unforeseen challenges, emphasizing real-time adaptation and recovery. In 5G, resilience was closely tied to security and trustworthiness, ensuring dependable operations through robust mechanisms aimed at mitigating vulnerabilities. In the context of 5G evolution, resilience extends beyond these aspects, becoming a core design principle embedded into link- and system-level frameworks. It now reflects the network's ability to adapt and reorganize dynamically, supporting the complex, intelligent demands of differentiated QoS, QoE, and AI-native applications, ensuring continuity across increasingly interconnected and autonomous systems. As a system goal, resilience is more apropos for 6G systems than reliability and robustness, because $(a)$ failures are inevitable and $(b)$ resilience assesses both long-term and short-term network performance, as compared to the short-term perspective of robustness. A resilient 6G network is designed to be adaptable, robust, and able to learn from experiences, ultimately ensuring continuous operation and minimal impact on users even in challenging or adverse conditions. Hence, a flexible network operation, evolution and learning, and an overall holistic system view become important.

	Along with resilience, in 6G networks and beyond, the criticality of applications becomes a crucial factor to be considered in network management \cite{9678373,wang2021smart}.
	In general terms, criticality refers to the degree of importance or significance of a system component, particularly in terms of its assurance against failure \cite{burns2013mixed}. Hence, a critical component must operate without malfunction to prevent severe consequences, e.g., human endangerment or environmental damage. The criticality level of 6G applications is typically determined by the potential risks and consequences associated with its failure. For example, safety-critical applications, e.g., in aerospace and healthcare, have failure consequences that are more severe, and, thus, they require greater assurances against failure. 
	The need to account for criticality becomes more pronounced when the network must deal with \emph{mixed criticality} \cite{7527646} scenarios, in which multiple services with different levels of criticality coexist.
    The related concept of \emph{differentiated connectivity} focuses on delivering varying levels of QoS tailored to the specific needs of users \cite{erricson}. In contrast, mixed criticality is application-focused, addressing the coexistence of critical and non-critical tasks and ensuring that critical tasks are given the necessary priority and protection. Differentiated connectivity serves as a key enabler for mixed-criticality systems.
    Industrial automation exemplifies the important role of mixed service criticality in 6G communication systems, as in this context, ensuring worker safety and environmental protection remain paramount even in the face of communication disruptions, while other services such as remote assistance calls are of lower priority. 
	This becomes particularly relevant when the risk and existing hazards of the environment increase. As a result, resilience must go hand in hand with criticality-awareness and appropriate service prioritization to efficiently restore functionality in accordance with individual application requirements and safety aspects despite resource limitations. 

    The recent work \cite{mahmood2024resilient} introduces the \emph{resilient-by-design} framework for 6G. By categorizing disruption types, defining resilience in wireless networks, and outlining a four-step approach to achieving resilience, it provides an excellent starting point for understanding resilience in 6G systems. However, the work does not delve into layer-specific implementation strategies or extensively addresses the integration of mixed criticality beyond application-level awareness, leaving significant gaps that we aim to address.
    
    The work in \cite{10716652} identifies three core principles for designing resilient 6G networks: $(1)$ integrating protective measures to address challenges during normal operation, $(2)$ incorporating self-awareness for failure detection, prediction, and performance evaluation, and $(3)$ enabling reconfiguration to adapt to challenges, learn from failures, and improve resilience policies. The study also emphasizes the need for an end-to-end system design that balances resilience with complexity and cost considerations. While \cite{10716652} offers a more detailed perspective on resilience across different network layers, the proposed layer definitions deviate from standard-compliant structures. Furthermore, the work does not address mixed-criticality scenarios, lacks in-depth discussion of individual protocol layers, and omits considerations related to the Open Radio Access Network (O-RAN) Alliance visions. 

    \subsection{Contribution}
	To this end, the main contribution of this paper is to pioneer the first comprehensive tutorial on integrating resilience and criticality-awareness into 6G network design. This requires defining how one can induce fundamental paradigm shifts from robustness and reliability to resilience and from service differentiation to criticality. The primary goal is to provide a stepping stone for a suite of solutions, that ultimately ensure the practical viability and safety of 6G-enabled critical applications, showcasing our vision of \emph{resilience and criticality} as \emph{brothers in arms for 6G}. 
 
    To achieve this, the paper presents guidelines for designing 6G networks that tightly integrate notions of resilience and criticality-awareness. First, we particularly explore the evolutionary leap from 5G to 6G, emphasizing the need to enhance robust network designs with resilience strategies. We propose categorizing services by their criticality to address mixed-criticality systems and discuss the unique challenges and opportunities of embedding criticality-aware resilience in 6G networks. Furthermore, we formally define resilience and criticality, present a core resilience strategy comprising three resilience cycles (robustness, short-term, and long-term resilience), and define a quantitative resilience metric. We identify leverage points across communication layers of 5G and O-RAN Alliance considerations. Resilience and criticality at the higher layers of 6G are supported in the control and management plane through network-wide monitoring, AI capabilities, and criticality-aware resource management. Virtualization of network functions enhances flexibility, empowering network slicing for failure isolation and rapid recovery. At lower communication layers, we address failure detection on the radio link, criticality-aware scheduling and access, and dynamic physical layer concepts. Finally, we provide an outlook on open research challenges, including artificial general intelligence, semantic communications, and scalability, paving the way towards future resilient and criticality-aware 6G systems.
    
    \subsection{Organization}
    The rest of this paper is organized as follows. Section~\ref{sec:6Gvision} presents the 6G vision on resilience and mixed criticality alongside its unique challenges and opportunities. A formal definition of resilience, a core strategy, and an evaluation metric are given in Section~\ref{sec:6Gresilience}. Then, Sections~\ref{sec:controlandmanagement}, \ref{sec:cloudvirtual}, and \ref{sec:lowlayer} review integrated 6G implementations of resilience and criticality-awareness on the control and management plane, as virtualization and slicing concepts, and on the lower communication layers, respectively. Section~\ref{sec:outlook} discusses outlooks and perspectives, as well as research challenges towards future resilient and criticality-aware 6G systems. Finally, Section~\ref{sec:con} concludes this article.
    
    A list of all acronyms used throughout this paper can be found in Table~\ref{tab:acronyms}.
    \begin{table}[]
        \centering
        \begin{tabular}{l p{6.5cm}}
        \hline \hline 
            3GPP & 3rd generation partnership project\\
            AI & artificial intelligence \\
            API & application programming interfaces\\
            A$^3$RT & anticipation, absorption, adaption, and recovery over time\\
            AGI & artificial general intelligence\\ 
            AoI & age of information\\
            BS & base station\\
            CP & control plane\\
            CPS & cyber-physical system \\
            CU & central unit\\
            DT & digital twin\\
            DU & distributed unit\\
            eMBB & enhanced mobile broadband\\
            ITU & international telecommunication union\\
            LoS & line-of-sight\\
            MAC & medium access control\\
            MIMO & multiple-input multiple-output\\
            ML & machine learning\\
            mMTC & massive machine-type communication\\
            NOMA & non-orthogonal multiple access\\
            NTN & non-terrestrial networks\\
            O-RAN & open radio access network \\
            OFDM & orthogonal frequency-division multiplexing\\
            PDCP & packet data convergence protocol\\
            PHY & physical layer\\
            QoE & quality of experience \\
            QoS & quality of service \\
            RAN & radio access network\\
            RIC & RAN intelligent controller\\
            RIS & reconfigurable intelligent surface\\
            RLC & radio link control\\
            RRC & radio resource control\\
            RF & radio frequency\\
            RSMA & rate-splitting multiple access\\
            RT & real-time\\
            RU & radio unit\\
            SDAP & service data adaptation protocol\\
            SMO & service management and orchestration\\
            UAV & uncrewed aerial vehicle\\
            UP & user plane\\
            URLLC & ultra-reliable and low-latency communication\\
            V2X & vehicular-to-everything \\
            VNF & virtual network function\\
            XR & extended reality \\
        \hline \hline 
        \end{tabular}\vspace*{.1cm}
        \caption{List of acronyms.}
        \label{tab:acronyms}
    \end{table}
	
    \section{6G Vision on Resilience and Mixed Criticality}\label{sec:6Gvision}
    In its origins, the term resilience roots back to the Latin verb \emph{resiliere}, meaning to rebound or recoil, e.g., "\emph{[...] saepe super ripam stagni consistere, saepe in gelidos resilire lacus [...]}", taken from ancient literature about a metamorphosis from human to frogs, which often sit at waterside and hastily jump back (recoil) into the cold water.  
    With roots in psychology, environmental studies, and biology, resilience and mixed criticality concepts bear significant promise for engineering domains, including wireless communication and cyber-physical systems (CPS). In CPS, physical entities are overseen and managed through a central hub of communication and computation \cite{Cassottana2023}. These systems, which integrate physical processes with computational control, require robust communication networks to ensure seamless operation and real-time responsiveness. CPS are prone to correlated failures because they have interdependent energy, computing, and communication resources, as well as complex and heterogeneous components, e.g., in the context of gas-power-water infrastructures \cite{8088670} or power grids \cite{7852237}. 
    Further, \cite{7852237} emphasizes the need for criticality-awareness in the face of emergencies, e.g., large-scale blackouts, when smart grids or emergency power generators must ensure operations of critical infrastructure such as hospitals, fire departments, police stations, etc. 
    Remarkably, despite this progress in resilience within fields such as CPS, those concepts are yet to make their way into wireless communication systems.

\subsection{From Robustness to Resilience: A Paradigm Shift}
Upon closer examination, the convergence of CPS and wireless communication systems becomes more evident. In CPS, the physical entities rely on continuous and reliable communication with the central hub to perform critical functions. Similarly, wireless communication systems, such as 6G, exhibit interdependent components, including network management, as well as core and RAN. Each of these components is highly complex, requires real-time coordination, and is mutually dependent, necessitating resilience for each part. Resilient communication, particularly via 6G networks, is therefore a cornerstone for the effective management and operation of CPS, ensuring that even in the face of unexpected challenges, the system can maintain functionality and quickly recover from disruptions.

However, to date, wireless networks have been primarily focused on network robustness, i.e., to withstand challenges and disruptions. While resilience has been considered in specific contexts, such as security frameworks and as an add-on feature, e.g., \cite{6584933,8029548}, it has not been a central design principle in previous wireless generations. Resilience in earlier systems was often limited to isolated use cases rather than being holistically integrated into the network architecture. For instance, 4G includes interference management, adaptive modulation, and access control, to mitigate impacts of various types of interference, dynamically adjust the modulation scheme, and prevent unauthorized access, respectively. However, 4G has severe single-points-of-failure in terms of radio resource control and cell association, e.g., losing a primary cell leads to a radio link failure \cite{robust5G}. Specifically, when faced with failures, the 4G network follows
a \emph{fail fast, recover fast} restarting principle \cite{robust5G}. 
The 5G standard significantly extends the robustness prospects in terms of cloud-native frameworks, redundant network functions, load balancing, and congestion control. Meanwhile, 5G includes a time-intensive recovery mechanism based on a user plane restoration procedure \cite{robust5G}, and includes centralization, leading to greater failure impacts. For critical services, 5G provides an ultra-reliable dual connectivity mode by establishing two redundant sessions. However, this still leaves the failure points in 5G RAN unaddressed. Moreover, much of the 5G design is reliability-centric \cite{7041045}, with a focus on ultra-reliable communications, such as in vehicle-to-everything (V2X) services \cite{8823846}.\looseness-1

Restarting mechanisms and the \emph{fail fast, recover fast} principle are only suitable for a subset of failures, specifically excluding unexpected and external disruptions. 5G RANs' single-points-of-failure constitute a significant vulnerability. Utilizing redundancies enhances the network robustness only up to some extend, specifically requiring the availability of similar homogeneous hardware/software components \cite{sankaran2022la}. Hence, a sole focus on reliability and robustness, as well as simple adaption and recovery mechanisms are unsuited for the challenges of future communication systems.
Consequently, future 6G networks should witness a paradigm shift from robustness, i.e., tolerating a failure without majorly changing the system state, to resilience, i.e., maintaining service, sustaining an adversary, overcoming a failure, and learning for future defects by altering the system state. Thereby, as illustrated in Fig.~\ref{fig:fig_1_circles_v4}, the robustness capabilities are included in the resilience paradigm as one partial aspect, while the framework also addresses additional wide-ranging facets such as remediation, recovery, and refining. Hence, 4G restarting principles and 5G redundancies need to be extended upon by more sophisticated resilience mechanisms. Instead of relying on pre-configured robustness techniques, 6G networks switch to proactive and autonomous remediation mechanisms, including already existing 5G considerations. 
They aim to handle unexpected and unforeseeable failures rather than relying on anticipated fault models. Instead of focusing on human-centered network hardening, 6G networks involve intelligence-centered automated system state refinement. This shift allows for more adaptable and resilient network management. 

The above-established importance of resilience in 6G networks is closely tied to another crucial factor: The coexistence of services with varying levels of criticality. The criticality level of each service is determined by the potential risks and consequences associated with its failure. Therefore, we next aim to gain a deeper understanding of integrating service criticality into 6G network management.
    
    \subsection{From 5G to 6G: Service Differentiation and Criticality}
\begin{figure}[!t]
	\centering
	\includegraphics[width=.8\linewidth]{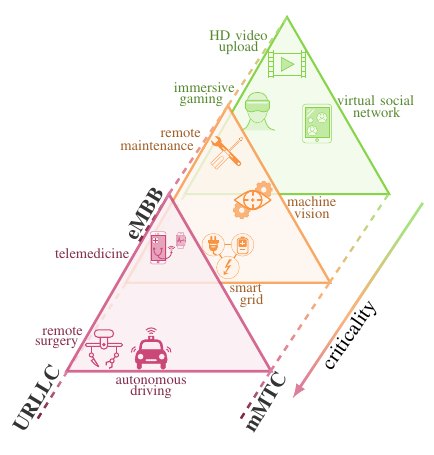}
	\caption{Use cases of 6G networks considering URLLC, eMBB, and mMTC network slices overlaid by criticality-awareness.}
	\label{fig:hexagon}
\end{figure}
Current 5G systems support three generic service classifications. According to ITU, such classes are referred to as enhanced mobile broadband (eMBB), ultra-reliable and low-latency communications (URLLC), and massive machine-type communications (mMTC) \cite{series2017minimum}. 
In short, eMBB requires stability and high peak data rates, URLLC represents a category reliant on reliable service and low latencies, and mMTC requires extreme coverage and a multitude of device connections.
In this context, URLLC already captures an early version of a critical communication service, e.g., alarm services or safety sensors might reside in this category triggering communication in emergency situations.
Hence, 5G includes the possibility of enabling the RAN to jointly serve co-existing functions of heterogeneous QoS requirements and of mixed-critical nature, e.g., by means of network slicing \cite{8476595}.

However, while current 5G networks already address both QoS and QoE requirements \cite{7224722}, 6G necessitates a more comprehensive approach to service differentiation. In the 6G vision, QoS is not the only factor to be considered, but is complemented by the inclusion of service criticality as an additional dimension, as can be seen in Fig.~\ref{fig:hexagon}. Indeed, services with similar demands on data rate or latency, which are sharing the same network slice, can considerably differ in their criticality levels depending on the use cases. For instance, an XR application for remote assistance purposes is much more critical compared to XR used for gaming or entertainment, and hence should be prioritized. Consequently, the classification of different service criticalities should not be restricted to QoS demands, but also take into account the importance of stringent compliance with those QoS requirements, i.e., how detrimental temporary QoS violation is for users and the environment. Fig.~\ref{fig:hexagon} presents an extended service classification, where the proposed 5G service triangle is overlayed by different criticality levels, showcasing a range of example use cases. For instance, eMBB includes critical telemedicine and uncritical high-definition video uploads.
During unexpected failures, these criticality levels are used to determine appropriate prioritization schemes among services. For example, within the same network slice, autonomous cars are given priority over social network users, and telemedicine services are prioritized above immersive gaming applications.

Both high-risk, critical services have more severe failure consequences, potentially endangering human lives. Therefore, resilience and criticality must go hand in hand when designing future 6G networks. We next examine the unique challenges and opportunities that 6G networks present for integrating criticality-awareness and resilience.

\subsection{Resilience in 6G: Unique Challenges and Opportunities}
With regard to the discussed factors, future 6G networks demand a novel and distinct perspective on \emph{resilience}. 

Firstly, going beyond ensuring robustness for expected errors, there is a need to incorporate flexible adaptation and recovery mechanisms to effectively address unforeseen challenges. In dynamic 6G environments, adaptation enables real-time anomaly detection and operational adjustments, while rapid recovery is vital for restoring functionality with minimal user impact. Specifically within 6G, this involves leveraging adaptable software components such as dynamically reconfigurable network functions, utilizing dynamic environment control through, e.g., reconfigurable intelligent surfaces (RIS), and the integration of AI and machine learning (ML) for enhanced network autonomy. 

Secondly, transitioning from a service-oriented perspective that prioritizes QoS, 6G demands a criticality-aware strategy that differentiates mission- and safety-critical applications demanding stringent dependability, ultra-low latency, and high availability. This shift emphasizes the need for dynamic resource allocation, leveraging advanced technologies like AI, and ensuring the prioritization and reliability of particular 6G applications like autonomous driving, V2X, and telemedicine.

Thirdly, for the assurance of overall system integrity, resilience must be seamlessly integrated into every layer and component of the network, demanding a holistic view of the network architecture. This means resilience strategies must be integrated across core network elements, edge devices, communication protocols, and management systems. A comprehensive framework covering all layers (hardware, software, control and user planes) enables 6G networks to mitigate risks and maintain stability amid component failures or disruptions.

Apart from that, the unique features envisioned for 6G present both challenges and opportunities concerning network resilience.
For instance, the work \cite{9040264} discusses 6G use cases and technologies, highlighting safety-critical applications that demand high QoS levels and emphasizing the need for autonomous, ultra-fast failure recovery. For example, V2X applications, such as collision avoidance and traffic management, require real-time communication, robust and low-latency responses, and rapid recovery to ensure safety and maintain smooth operation.  
Moreover, the decentralized nature of 6G networks offers opportunities for resilience, efficiency, and low-latency communication. While it mitigates 5G RANs' single-points-of-failure, coordination and resource management pose significant challenges.
In addition, starting from 5G, network virtualization and software-defined networking provide flexibility, adaptability, and scalability, facilitating self-healing and reconfiguration. Nevertheless, they pose significant challenges in terms of complexity and reliability, particularly as both soft- and hardware failures are of equal importance.

Numerous 6G technologies offer improvements in high data rates, reliability, and sustainability, but they also introduce new vulnerabilities and challenges. For instance, utilizing higher frequency bands increases susceptibility to signal attenuation and blockages. Likewise, the interdependency of services, as well as the convergence of communication, sensing, control, localization, and computing in 6G networks can exacerbate the impact of communication disruptions \cite{saad20206G}. 
To categorize various types of disruptions, the \emph{resilient-by-design} framework proposed in \cite{mahmood2024resilient} classifies them into internal, unintentional external, and cybersecurity-related factors. Internal factors encompass challenges such as wireless channel impairments and software bugs, while unintentional external factors include disruptions like natural disasters. Cybersecurity-related factors, on the other hand, refer to deliberate malicious threats, such as cyber-attacks, aimed at compromising network integrity.
Finally, a resilient 6G system must consider the limited availability of resources, particularly in light of the pressing demand for energy efficiency \cite{nextGAroadmap}. Consequently, criticality-awareness and prioritization schemes emerge as crucial aspects for resilient 6G system designs.

\section{Formal Definition of Resilience}\label{sec:6Gresilience}
Consider a scenario in autonomous driving where resilience is critical across all network layers. At the air interface, resilience involves maintaining robust communication despite challenges like fading, interference, or blockages. For instance, a truck blocking the line of sight in a millimeter-wave or THz communication link can disrupt connectivity. Without resilient mechanisms, this disruption could result in the loss of critical control data necessary for vehicle coordination. At the MAC layer, resilience ensures timely scheduling of critical data, such as obstacle detection or braking commands. A failure here might occur if resource contention causes delays in high-priority safety messages while infotainment traffic consumes channel resources. Such delays could lead to dangerous outcomes, including potential collisions. At the network layer, resilience relies on route redundancy to maintain reliable data delivery. If a local BS fails, the vehicle might lose its connection to the core network, disrupting real-time processing of traffic updates or route optimization, potentially compromising safety and efficiency. At the application layer, resilience is essential for failover mechanisms that keep safety-critical applications operational. A disruption at this layer, such as an AI model on the cloud failing to provide timely pedestrian detection updates due to incomplete data, could result in missed obstacle detection and increase the risk of accidents. Failures in resilience at any layer can cascade through the system. For example, a disruption at the air interface may lead to delays in MAC-layer scheduling, network-layer routing failures, and ultimately, the inability of the application layer to make timely safety-critical decisions. Such a breakdown could result in life-threatening situations, such as a vehicle collision.

Alongside such examples, the evolutionary leap from 5G to 6G and the paradigm shift towards resilience and criticality-awareness establish the groundwork for adopting a unified resilience strategy in the context of 6G. 
Resilience strategies, which were for example designed for 4G, 5G, and various CPS, are not suitable to directly apply to 6G networks. For instance, relying on static redundancy is rather optimistic and costly in a large-scale, widespread communication infrastructure like 6G. Also, component restoration based on human intervention does not fit the time-scale of many 6G use cases, which require low-latency and ultra-reliability, e.g., XR, autonomous driving, and digital twins \cite{chaccour2023joint}. Other resilience strategies, which were designed for CPS \cite{Cassottana2023}, could potentially be relevant but they must be extended and adopted to the challenges in the context of 6G networks.
There is a plethora of promising new technologies envisioned for 6G that have immense potential to serve as key enablers of criticality-aware resilience; 
examples include integrated sensing, AI and ML, and network slicing \cite{nextGAroadmap}.
Yet, effectively leveraging those technologies requires a comprehensive framework for resilient network design. Hence, we next present a core resilience strategy, a promising quantitative resilience metric, and design criteria for resilience.

\subsection{Core Resilience Strategy}
\begin{figure}[!t]
	\centering
	\includegraphics[width=.5\linewidth]{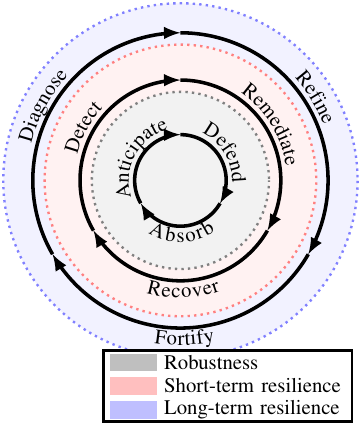}
	\caption{Sketch of the proposed resilience strategy relying on robustness, short-term resilience, and long-term resilience. Each cycle is empowered by three unique phases of the resilience process.}
	\label{fig:resilience_cycle}
\end{figure}
An illustration of the proposed core resilience strategy can be seen in Fig.~\ref{fig:resilience_cycle}. The inner-layer (first-order resilience \cite{jaeger2010risk}; core-defense \cite{STERBENZ20101245}; prediction and preemtion \cite{mahmood2024resilient}) is referred to as \emph{robustness}. For robustness, the network is defended against expected error configurations, anticipated adversaries, and known risks, the influences of which should be absorbed by the defense mechanisms. Robustness classifies into \emph{passive}, e.g., tolerating channel state uncertainties \cite{10066838}, and \emph{active}, e.g., 5G mobility robustness optimization \cite{9770833}, depending, however, on prior knowledge of possible error configurations. Thereby, robustness consists of three key phases: \emph{anticipate}, \emph{defend}, and \emph{absorb}, which involve redundancy and diversity, e.g., via multiple antennas, hybrid frequency utilization, or dual connectivity.

The middle-layer (second-order resilience \cite{jaeger2010risk}; D$^2$R$^2$ inner-loop \cite{STERBENZ20101245}; reactive recovery \cite{Cassottana2023}; protection \cite{mahmood2024resilient}) is referred to as \emph{short-term resilience}. This resilience cycle becomes vital when the network robustness methods fail, e.g., in unexpected situations. After detecting the failure, e.g., by using QoS deviation measures \cite{comebackkid}, the network controller employs different remediation mechanisms in order to recover the functionality to a reasonable level. Short-term resilience consists of \emph{detect}, \emph{remediate}, and \emph{recover} phases, which involve resource management solutions \cite{10104574}, adaption mechanisms \cite{comebackkid}, criticality prioritization \cite{comebackkid}, and knowledge-based autonomic behavior \cite{1160055}.

The outer-layer (third-order resilience \cite{jaeger2010risk}; DR outer-loop \cite{STERBENZ20101245}; proactive long-term planning \cite{Cassottana2023}; progression \cite{mahmood2024resilient}) is referred to as \emph{long-term resilience}. After a successful iteration of the short-term resilience, the long-term resilience aims at evolving the network operations. Diagnosing the recent event, the network can be refined by modifying system functions, resource allocation, or introducing further defense mechanisms to fortify the overall network operation in the long-term. Long-term resilience thereby involves \emph{diagnose}, \emph{refine}, and \emph{fortify} phases, which are traditionally based on human interaction, i.e., operators manually diagnose, refine, and fortify the network operations. In 6G, this process is envisioned to be sped up by dynamic architectures and protocols, AI, as well as self-optimization, self-healing, and self-protection mechanisms \cite{1160055}.
Such a constantly evolving resilience policy requires continuous evaluation and analysis of the network's resilience behavior, highlighting the need for a unified and comprehensive resilience metric.

\subsection{Quantitative Evaluation of Network Resilience}
\begin{figure}[!t]
		\centering
		\includegraphics[width=.95\linewidth]{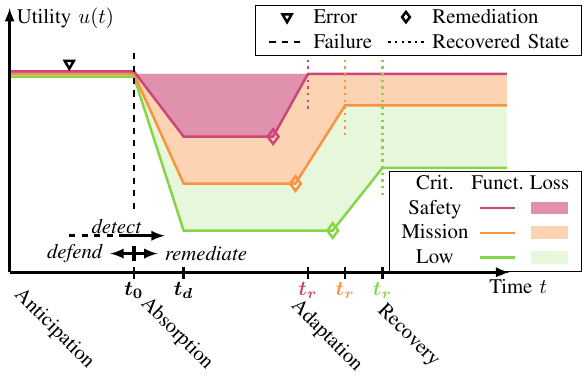}
		\caption{The A$^3$RT-chart: Illustrative resilience behavior of a mixed-critical functionality communication system.}
		\label{fig:resilience_1}\vspace*{-.1cm}
	\end{figure}%
\begin{figure*}[t]
	\begin{align}\label{eq:en}
	    \text{resilience} = \lambda_1 \underbrace{\frac{ \int_{t_0}^{t_d} u(t) \, dt}{\int_{t_0}^{t_d} u^\mathrm{des}(t) \, dt}}_{(i)} + \lambda_2 \underbrace{\frac{ \int_{t_d}^{t_r} u(t) \, dt}{\int_{t_d}^{t_r} u^\mathrm{des}(t) \, dt}}_{(ii)} + \lambda_3 \underbrace{t_\mathrm{rec}}_{(iii)} &&t_\mathrm{rec} = \begin{cases}
			1 \quad\quad\;\; t_r-t_0 \leq \Delta t_r^\mathrm{des}\\
			\frac{\Delta t_r^\mathrm{des}}{t_r-t_0} \quad \text{otherwise}
		\end{cases}
	\end{align}
	\hrulefill
\end{figure*}%
    A network achieves resilience if and only if each subsystem, whether hardware or software, is itself resilient. This principle applies across individual network features, components, and protocol layers. From a layer perspective, each layer must independently ensure functionality under adverse conditions, while cross-layer integration enables the coordination needed to maintain overall network performance. To measure and quantify the efficacy of network resilience, whether at the subsystem level or for the entire network, we next introduce a resilience assessment metric consisting of \emph{absorption}, i.e., the capability of the service to maintain the functionality in case of a failure, \emph{adaption}, i.e., the quality of the remediation mechanism utilizing the network resources to tackle the adversary, and \emph{time-to-recovery}, i.e., the time until a stable state (preferably normal operational state) is reached. As a note, the \emph{time-to-recovery} is proportional to the perceived \emph{time-to-content}, i.e., the responsiveness of a service. Critical applications require both minimal time-to-content and near-instant recovery, while non-critical services can tolerate longer recovery periods and slower responsiveness. With the emergence of AI-native devices and UEs, which are increasingly self-sufficient in computation, the \emph{time-to-recovery} becomes increasingly dependent on network availability, local and distributed compute capabilities, as well as queuing delays between devices and the network. 
	Fig.~\ref{fig:resilience_1} illustrates the core concept of criticality-aware resilience, providing a chart as a visual representation of the fundamental principles to system functionality and operation. In particular, Fig.~\ref{fig:resilience_1} plots a generic system utility $u(t)$ (e.g., in 6G, QoS or user experience) over a time axis for safety-, mission-, and low-critical network services. For instance, such services could map to particular 6G use cases such as autonomous driving, remote maintenance, and XR gaming, respectively. At the left-hand side of the plot, the available resources are sufficient for the 6G network to provide satisfactory functionality for all services. After an error, e.g., hardware failures, blockage, or attacks, we observe decreasing functionalities, called failure. In 6G, different precaution techniques such as multi-connectivity and robust beamforming may lead to a less severe functionality drop of critical services. Remediation mechanisms, such as utilizing redundant paths, beam adaptation, or frequency fallbacks, are able to (partially) recover the network functionality. In this context, resilience entails different phases, i.e., \emph{defending}, \emph{detecting}, \emph{remediating}, and metrics (see also \cite{9963527,Najarian2019}), i.e., \emph{Anticipation}, \emph{Absorption}, \emph{Adaptation}, and \emph{Recovery over Time}, leading to the A$^3$RT-chart in Fig.~\ref{fig:resilience_1}.

    To be more specific, Fig.~\ref{fig:resilience_1} entails three relevant time points namely, the time of failure $t_0$, the time of the lowest utility $t_d$, and the time of recovery $t_r$. Note that $t_d$ can also be referred to as the time when a stabilized utility is reached after failure. Let an arbitrary network utility be denoted as $u(t)$ and the desired utility as $u^\mathrm{des}(t)$, both of which are time-dependent. The work in \cite{Najarian2019} classifies such utility functions $u(t)$ to capture system performance, ensuring they are easy to implement, measurable, or at least computable from measurable values to reflect the current system state. Multiple time points with their corresponding utility values are required, specifically at $t_0$, $t_d$, and $t_r$. For instance, \cite{comebackkid} introduces a utility function based on the sum rate of users, while \cite{shui2024design} employs the time to collision in a V2X environment. 
    A mathematical representation of the A$^3$RT resilience metric can be found in \eqref{eq:en} (on top of the next page), where $\Delta t_r^\mathrm{des}$ is the desired time-to-recovery. More specifically, \eqref{eq:en} entails the absorption $(i)$, adaption $(ii)$, and time-to-recovery $(iii)$ components, and the corresponding weights $\lambda_i$, $i=1,2,3$, with $\sum_{i=1}^3\lambda_i = 1$ \cite{Najarian2019}. Related resilience definitions in the literature often focus on just one component, such as absorption, adaptation, or time-to-recovery, or tend to be heavily biased toward one aspect. A general metric like \eqref{eq:en} offers the advantages of being adaptable to operator needs and sufficiently broad to accommodate various utility functions, such as sum rate or time to collision. This metric can also be extended to numerically account for pre-failure anticipation. For a more comprehensive discussion on resilience metrics, we refer to \cite{Najarian2019}.
    
    We note that Fig.~\ref{fig:resilience_1} only shows an example resilience behavior for three different criticalities within the network. The exact A$^3$RT-chart behavior is highly network and situation dependent, e.g., the low-critical functionality could be completely halted for some time, or the adaption could already start during the absorption phase.
    Another note, the design of the resilience strategy and A$^3$RT-chart behavior should be based on an analysis of service criticality, the risk of a particular service, and the consequences of its complete ruin. 
    For instance, in case of a high risk of ruin, anticipation/defending (proactive) and absorption are most relevant, whereas the focus for rare or less harmful errors is rather on remediation/recovery (reactive) \cite{Cassottana2023}. In \eqref{eq:en}, the weights $\lambda_1$, $\lambda_2$, and $\lambda_3$ could, therefore, be tuned to achieve a desired outcome. 

    To achieve high levels of resilience, the 6G network design must proactively incorporate various criteria to integrate resilience capabilities, which we explore next.

\subsection{Design Criteria for Resilience Implementation}
Given the formal definition of resilience and its quantitative evaluation metric \eqref{eq:en}, it is crucial that resilience strategies address all cycles (Fig. \ref{fig:resilience_cycle}) and phases (Fig. \ref{fig:resilience_1}) to be effective. To accomplish this, 6G network design must integrate the following key criteria across various layers and components of the network.

Following the cycles outlined in Fig.~\ref{fig:resilience_cycle}, the upper-left phases of all cycles, namely the anticipation, detection, and diagnosis phases, are all dependent on system observation and monitoring. Furthermore, to calculate the resilience metric \eqref{eq:en}, the utility $u(t)$ has to be measured. Therefore, continuous \emph{system state monitoring} is a fundamental element of resilient network design.
Monitoring can occur at three time scales: real-time (RT), near-RT, and non-RT. RT monitoring (at around $1$ms) is crucial for functionalities of the lower communication layers, such as scheduling. Moving beyond the RT scale, RAN control functionalities in the near-RT and non-RT scales manage broader monitoring and control tasks. In the near-RT scale ($10-100$ms), short-term radio resource management is performed, while non-RT systems ($\gg 1000$ms) focus on long-term RAN setup and network organization \cite{5gbook}. For example, the O-RAN Alliance introduces RAN Intelligent Controllers (RICs) in this context, specifically the Near-RT RIC and Non-RT RIC \cite{oran_architecture}. Comprehensive monitoring approaches are essential for error prediction and anomaly detection, including channel prediction and traffic variation monitoring. Furthermore, criticality-aware monitoring involves a careful assessment of risks, their impact on mixed-critical applications, and the evaluation of safety requirements. 

Subsequently, within the robustness cycle, the network defends against errors and absorbs potential failure impacts. The goal is to minimize the utility drop after the time of failure $t_0$, especially for safety-critical services, as illustrated in Fig.~\ref{fig:resilience_1}. To achieve this, integrating \emph{redundancy and diversity} in both software and hardware is essential. In 6G, redundancy is inherently integrated into the network design. This includes envisioned support for the management plane, RAN controllers, and the cloud-based protocols of the central unit (CU) and distributed unit (DU) \cite{5gbook}. For instance, O-RAN specifies an O-Cloud to facilitate such redundancy \cite{oran_architecture}. This cloud-native RAN implementation enables the use of redundant cloud resources, enhancing interoperability, scalability, and resilience against unpredictable failures. On the lower layers, resilience measures include redundancy through error-correction codes, multi-connectivity, and multi-path diversity enabled by technologies like RIS. Recognizing the mixed-critical nature of applications, different levels of redundancy and diversity can be applied based on criticality levels, balancing the trade-off between the costs of redundancy and the importance of the application. This ensures that highly critical applications receive more robust redundancy measures, while less critical services are managed with cost-effective solutions.

For remediation and recovery in Fig.~\ref{fig:resilience_cycle}'s short-term resilience, as well as for the long-term refinement of resilience policies and network fortification, the \emph{adaptability and flexibility} of protocols and infrastructure are crucial. This approach spans both the adaptation and recovery phases depicted in Fig.~\ref{fig:resilience_1}. In 6G, these capabilities are significantly enhanced by the near-RT and non-RT RAN control functionalities, which oversee control tasks, resource allocation, RAN setup, and network organization.
In the near-RT scale, swift resource reallocation prioritizes critical applications during failures, enabling fast recovery. Meanwhile,  non-RT control enhances long-term resilience by adjusting access schemes and revising radio unit (RU) allocation policies to fortify the network, ensuring sustained robustness against future failures.
For example, in O-RAN, the Non-RT RIC operates within the broader Service Management and Orchestration (SMO) framework that covers administrative tasks, policy management, and AI integration. By leveraging AI, resilience strategies employ \emph{intelligence} across all phases, enabling prediction, self-healing, and autonomy.

Given the complexity of the 6G architecture, coupled with decentralized network organization and interdependencies among various components and protocol layers, a comprehensive approach is necessary to achieve resilience. This requires addressing resilience challenges across the entire protocol stack. The following sections explore criticality-aware resilience techniques at different communication layers: Control and management (Section~\ref{sec:controlandmanagement}), virtualization and slicing (Section~\ref{sec:cloudvirtual}), and resource allocation, as well as signal and packet processing at the lower layers (Section~\ref{sec:lowlayer}).

\section{Control and Management Plane}\label{sec:controlandmanagement}
\begin{figure}[!t]
	\centering
	\includegraphics[width=\linewidth]{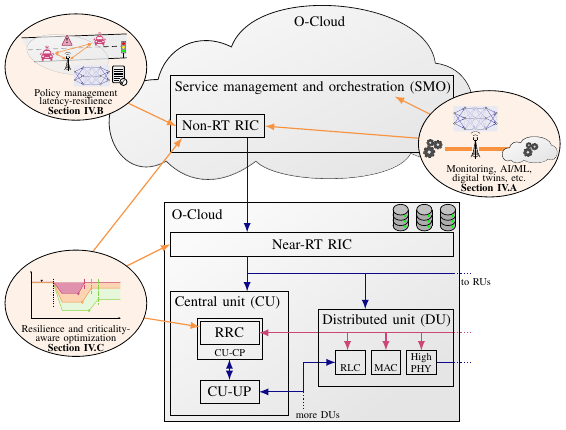}
	\caption{Management and control plane snippet of the 5G/O-RAN protocol stack showing technologies for resilient and criticality-aware network operation.}
	\label{fig:resilience_CPandManage_v2}
\end{figure}
While designing resilient communication systems requires a comprehensive approach spanning various layers of the protocol stack, the effective coordination of resilience strategies and criticality-awareness is paramount. In particular, short- and long-term resilience, as well as system state monitoring, fall within the scope of the central coordination implemented on the management and control plane. Serving as the backbone of the overall resilience strategy, they are responsible for overall orchestration and decision-making of both proactive and reactive measures, ensuring seamless adaptation across all phases of resilience.
Therefore, Fig.~\ref{fig:resilience_CPandManage_v2} shows the management side with SMO and Non-RT RIC, and the control plane (CP) including Near-RT RIC and CU-CP. The CU-CP hereby consists of the radio resource control (RRC) layer and controls all subsequent RAN components, e.g., the DU. 

This section, in particular, briefly reviews the resilience prospects of the management plane in terms of monitoring, analysis, and AI integration, policy management in terms of latency-resilience trade-offs, as well as the notion of resilience and criticality-aware optimization approaches for network management and resource allocation on the CP.

\subsection{Management Plane -- SMO and RIC}
Considering resilience on the management plane, the focus is rather on the long-term perspective, as functionalities of the SMO reside in a non-RT ($\gg 1000$ms) control cycle, managing policies and the long-term RAN setup.
In the SMO block, resilience management involves network monitoring and analysis, formulation of response strategies and decision-making, and resilience evaluation within a continuous control cycle, e.g., see Fig.~\ref{fig:resilience_cycle}. 
Prospective SMO-based RAN resilience frameworks comprise two modules as part of the network management system, namely an analysis module and an estimator module \cite{9942669}. The analysis module learns from past failure logs to identify and model potential errors and vulnerabilities of the network and enable the prediction of failures, risks, and the impacts of different mitigation strategies. The estimator module is responsible for the selection and activation of suitable response mechanisms. Hence, the analysis and estimator modules map to the \emph{diagnose} and \emph{refine/fortify} phases of the long-term resilience in Fig.~\ref{fig:resilience_cycle}, respectively. In terms of costs, the framework in \cite{9942669} leverages existing network metrics, such as the 3GPP-defined reference signal received power, minimizing the extra expense of additional resilience analysis. However, estimating appropriate failure response mechanisms requires additional computing resources within the management and control plane.\looseness-1

A fundamental prerequisite for resilience management is an awareness of the current state and general dynamics of the network. 
To this end, \cite{9942669} proposes a coverage based resilience analysis using a Markov model with states ranging from good coverage to outage. The transition between states is determined by the occurrence of different degrees of perturbations on the one hand, and resilience mechanisms with varying efficacy on the other hand.
In addition to model-based approaches, AI becomes a key enabler for a precise network analysis, prediction, and smart adaptable decision policies. 
In contrast to model-driven approaches, which rely on predefined rules and structures (which are subject to frequent changes in 6G), data-driven approaches learn patterns, make predictions or decisions, and generate outputs for AI-aided network management. Thereby, such AI-based approaches can parse through large amounts of data, which particularly suits Fig.~\ref{fig:resilience_CPandManage_v2}'s SMO block, as it has an overview perspective and global knowledge of the RAN \cite{10329947}. In particular, an AI-aided SMO block can help facilitate network analysis in terms of prediction mechanisms, and smart adaptable decision policies. In more details, RAN elements pass different data, e.g., performance metrics or telemetry, to the SMO via different O-RAN interfaces. An AI-aided controller performs prediction of the RAN status or anticipates dynamic network behavior. Based on such predictions, reconfiguration mechanisms control the underlying RAN component by changing the policy, e.g., slice configuration or service provisioning \cite{10329947}. 

A particularly promising approach envisioned for 6G are network digital twins (DTs), i.e., digital representations of the physical network, which mirror its behavior and functionality within the environment \cite{10179151}. By continuously updating a virtual network model based on real-time data, DTs facilitate monitoring, predictive analysis, and optimization of the network. In particular, a DT can reflect traffic fluctuations, user mobility patterns, and predict errors, disruptions, or security attacks, thereby allowing for proactive mitigation or prevention of failures. Furthermore, by enhancing the detection of faults and anomalies, the recovery time can be reduced. 
Moreover, when combined with extensive sensor data from the environment, the Metaverse complements DT capabilities by representing the entire network and its surroundings in a comprehensive virtual environment \cite{9833928,10279406}. DT simulations and Metaverse representations can also generate training scenarios for AI-driven fault recovery systems, enabling timely and accurate response mechanisms at the RIC level.
In addition, the simulation of potential errors or changes in the network allows for a detailed risk analysis, and subsequently, simulating the impact of potential remediation strategies facilitates decision-making. Hence, AI-based network management and DTs play a key role in all resilience phases shown in Fig.~\ref{fig:resilience_1}.

\subsection{Policy Management: Latency vs. Resilience -- Non-RT RIC}
    There are further ranging functionalities in addition to the above described monitoring, analysis and prediction functionalities of the SMO block and, in particular, the Non-RT RIC. According to the O-RAN specifications, the Non-RT RIC is responsible for RAN policy optimization, model training for AI/ML, and supporting the underlying Near-RT RIC's operation in general \cite{oran_architecture}. Thereby, its primary goal is to achieve high-level non-RT objectives, e.g., long-term resilience as depicted in Fig.~\ref{fig:resilience_cycle}. 
    Through policy management, a guideline for the behavior of the network over a longer time scale is defined and enforced. Thereby, the Non-RT RIC governs how the network allocates and manages resources.     
    By that, the Non-RT RIC handles an important aspect, namely the interplay between latency and achievable resilience.

    In particular, 6G networks inherently support safety-critical applications like autonomous vehicles that require ultra-low latency, and, hence, have less time to react to failures or errors. This introduces a fundamental limit of the types of resilience strategies that can be employed. For instance, certain resilience mechanisms (e.g., rerouting traffic, switching to backup systems, or performing re-optimizations) introduce additional delays, e.g., as shown in \cite{comebackkid}. To maintain low latency and real-time performance, resilience mechanisms must be highly efficient and introduce minimal overhead. This often means reducing the complexity of remediation and recovery processes, as per Fig.~\ref{fig:resilience_1}, potentially sacrificing some degree of resilience or performance level after recovery.

    One solution approach is the introduction of AI/ML capabilities, as envisioned by the ITU in every part of the communication system \cite{IMT-2030}. In that sense, AI-driven decision-making, and ML inference boost the performance of certain resilience mechanisms. Instead of re-running an algorithm to solve an optimization problem for resource management (e.g., see \cite{comebackkid}), an ML model generates resource allocations in real time reacting to, for instance, channel or computing uncertainties in the network \cite{reifert2024robust}.
    Another solution approach, the mixed-critical nature of 6G networks requires resources to be allocated in a way that balances the needs of both high-latency-tolerant and low-latency-critical applications. This opens the opportunity of prioritizing certain traffic, which can lead to trade-offs in resilience for less critical services. Latency-critical services often demand a stronger emphasis on proactive strategies, such as redundancy or diversity, to ensure immediate protection and robustness. Conversely, less critical services, being more tolerant of adaptation delays, can rely on resource-efficient reactive strategies that activate only in response to failures \cite{karacora2024resilience}. A weighted resilience metric, such as \eqref{eq:en}, enables resource allocation that addresses the heterogeneous needs of 6G applications while ensuring minimal recovery times for critical services. Overall, the design of 6G networks, especially the Non-RT RIC must account for this trade-off, operating adaptively and in a context-aware fashion to determine policies for enabling a smooth interplay between latency and resilience according to the needs of different applications and services.
    
\subsection{Radio Resource Control -- RIC and CU-CP}
    According to 3GPP, the RRC as part of the CP is responsible for hosting a finite state machine for different operational states and transitions of the network. 
    According to the O-RAN specification, the Near-RT RIC is responsible for controlling and optimizing RAN functions in a near-RT control loop ($>10$ ms). As a result, the Near-RT RIC plays a crucial role in enhancing robustness and facilitating short-term resilience. 
    Recalling its definition, resilience is the ability of the network to change its configuration or state in order to overcome the failure. Hence, the Near-RT RIC poses a significant enabler of resilience for the RAN. It is specifically able to change states, state transitions, and the overall policy of the RRC, which hosts only a fixed state machine to control the RAN. Therefore, the RIC is augmenting the 6G adaptability and evolvability, particularly in terms of resilient and criticality-aware radio resource management, which we focus on next.
    
    The general goal of radio resource management is to allocate limited radio resources, e.g., time-slots, frequency spectrum, and power, for optimized performance and efficiency of the network. The goal of resilient resource management is twofold: On one hand, a solution should be as \emph{robust} (inner resilience cycle, Fig.~\ref{fig:resilience_cycle}) as possible, and on the other hand, should natively include adaption and remediation mechanisms for short-term resilience. Along such lines, criticality-aware resource management provides the necessary means to prioritize critical services, making the initial resource allocation more robust for such services, and, in the short-term, give precedence during failure-induced resource scarcity.

    Combining criticality awareness with resilient resource management enables high-quality and robust resource allocations during regular operation while ensuring that greater portions of available resources are assigned to critical users (see \eqref{eq:en} and the A$^3$RT-chart in Fig.~\ref{fig:resilience_1}). For instance, cloud-radio access networks require the management of parameters such as data rates, beamforming vectors, and user clustering, particularly by leveraging criticality-aware resilience strategies \cite{comebackkid}. Network robustness is further enhanced by multi-connectivity, including user-to-base station (BS) clustering and rate-splitting, which will be discussed in Sec.~\ref{sssec:noma}. In the event of a failure, \cite{comebackkid} suggests employing four resilience mechanisms, i.e., rate adaption, beamformer adaption, BS-user-allocation adaption, and comprehensive adaption, differing in calculation time and quality-of-recovery. Following the listing, such mechanisms depict resource re-allocations of increasing complexity, e.g., applying a simple formula for rate allocation, to a full-fledged re-optimization of resources. As main takeaways, the results in \cite{comebackkid} illustrate the superiority of resilient over non-resilient networks, the importance of criticality-awareness, and that multiple resilience mechanisms need to be combined for providing amenable performance and high resilience levels over a broad range of networks and failure parameters.

    Another crucial aspect of resilient radio resource management is the feasibility of resource allocation optimization problems. Prospective optimization algorithms must be adaptable to handle infeasibility under extreme network conditions \cite{10066838}.
    Due to the stochastic nature of the wireless channel, minimum-rate and QoS constraints can often become infeasible, leading to failed resource allocation solutions. By incorporating slack variables and penalty terms, a reformulated optimization problem ensures that the radio resource management algorithm consistently generates a solution, even in the face of unforeseeable channel conditions. 

    As an interim summary, 6G shows promising prospects for criticality-aware resilience on the control and management plane. Specifically, as shown in Fig.~\ref{fig:resilience_CPandManage_v2}, the integration of the AI-native SMO, coupled with the monitoring, analysis, and management capabilities of the RICs, along with criticality-aware resource management on the Near-RT RIC and RRC, holds potential to enhance the resilience of 6G networks.

    Several of the 6G resilience features discussed above depend on dedicated network functions, such as monitoring, prediction, and resource optimization, which themselves must be both flexible and resilient. Virtualization frameworks in 6G aim to deliver these capabilities by enabling scalable, dynamic, and reliable software-based network functions, often leveraging cloud-native principles. Building on this, upper-layer communication protocols facilitate advanced concepts like network slicing and recovery mechanisms that not only manage mixed-critical services but complement the overall network resilience. Resilience and criticality operate in tandem from an end-to-end perspective, spanning both the core network and the RAN. In the following section, we investigate the roles of virtualization, cloudification, and network slicing, emphasizing their importance for achieving resilience and criticality awareness in 6G networks, with a focus on core network functionalities.   

    \section{Virtualization and Slicing Frameworks}\label{sec:cloudvirtual}
    \begin{figure}[!t]
	\centering
	\includegraphics[width=\linewidth]{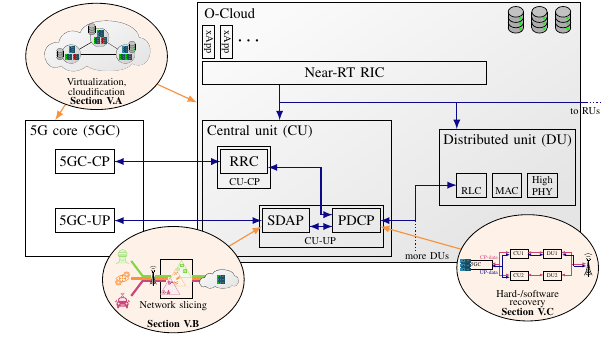}
	\caption{Opportunities for resilience and criticality-awareness on the higher layer 5G/O-RAN protocol stack, O-Cloud, 5G core, and CU.}
	\label{fig:resilience_CloudandVirtual}
    \end{figure}  
    General-purpose compute nodes and data centers are capable of hosting virtualized network functions, transforming tasks like policy management in the SMO, network slicing and session management in the CU, and scheduling in the DU into software-based processes. In fact, by abstracting hardware functionalities into software, virtualization lays the foundation for enhanced network flexibility, allowing for rapid scaling of services, adaptation to changing conditions, and dynamic prioritization. For instance, network slicing enables the segmentation and isolation of mixed-critical traffic flows, ensuring that each slice meets the required QoS for its criticality level. Additionally, isolation of potential failures and per-slice network configuration become possible, facilitating the tuning of resilience levels to match the slice-specific needs.
    
    Fig.~\ref{fig:resilience_CloudandVirtual} shows a high-level view of the potential 6G protocol stack for the abstract 5G core, the RAN, as well as the Near-RT RIC. The 5G core and the CU are particularly split into CP and UP. 
    On the RAN side, the CU-UP comprises the Service Data Adaptation Protocol (SDAP) and Packet Data Convergence Protocol (PDCP). The SDAP is responsible for managing QoS of different data flows, while the PDCP performs compression, encryption, and arranges packet orders, sequencing, and duplication. Fig.~\ref{fig:resilience_CloudandVirtual} additionally shows how a CU can connect to different DUs, and the Near-RT RIC runs different xApps, i.e., different software applications of network functions (virtual network fuctions (VNFs)).
    
    We next particularly review the notions of a virtualized, cloud-native RAN, consider network slicing via QoS flows, and recovery mechanisms on the PDCP layer. Thereby, we emphasize each technology's potential to empower resilient 6G networks that also account for mixed criticality.
    
    \subsection{Virtualization and Cloudification -- O-Cloud}    
    As a prerequisite, virtualization in the 6G context effectively combines softwarization and programmability. \emph{Softwarization} refers to the transformation of traditionally hardware-based network functions into fully software-implemented components, while \emph{programmability} enables these functions to be dynamically controlled, configured, and adjusted through specific interfaces. Together, these aspects allow network functions to operate in virtualized environments, enhancing the network's flexibility, scalability, and resilience. Furthermore, \emph{cloudification} involves migrating services and infrastructure to cloud-based environments using generic compute resources, thereby transforming systems into cloud-native architectures.
    In terms of the protocol stack shown in Fig.~\ref{fig:resilience_CloudandVirtual}, a cloud-native 6G network integrates software-driven RICs, CUs, and DUs. Unlike 4G, where monolithic BSs integrate all network functions, the O-RAN architecture inherently supports virtualization, cloudification, and disaggregation, as detailed in \cite{10329947}.
    As a note, while virtualization and VNFs are already key components of 5G networks, there remain unsolved technical issues in 5G \cite{9380157}. Consequently, 6G VNFs are envisioned to feature enhanced flexibility, AI-driven capabilities, deeper cloud-native integration, and criticality awareness. These advancements will make VNFs more dynamic, adaptive, and capable of supporting the more stringent demands of future networks, positioning them as essential elements in the paradigm shift towards greater resilience and criticality.

    Transforming the network architecture towards the aforementioned concepts is more than just an architectural shift. Rather, it is a critical step in enhancing the resilience and criticality awareness of 6G networks. Softwarization allows rapid deployment, scaling, and reconfiguration of network functions, now implemented as VNFs, in response to failures or changing network conditions. Focusing on robustness, VNFs can be deployed on general-purpose and off-the-shelf compute nodes, where a suitable virtualization layer enables shared backup resources to protect network services against physical network component failures \cite{7919487}. Both the network virtual links and VNF instances are protected by physical backup paths and nodes, respectively. Therefore, the algorithm in \cite{7919487} determines backup instances which are disjoint from the active links/nodes during regular operation. The states of VNFs should be managed by a centralized state manager, which oversees and triggers resilience phases as depicted in Fig.~\ref{fig:resilience_cycle} \cite{7981543}.
    Going beyond robustness, programmable elements dynamically adjust network configurations to prioritize resources for high-criticality applications in the event of a failure. As an example, at least two redundant active VNFs can be maintained for critical applications, as opposed to relying solely on spare standby resources for lower-criticality services \cite{7981543}. 
    Such a VNF-based approach facilitates the concept of \emph{modularization}. In modular communication architectures, application programming interfaces (APIs) enable seamless communication and coordination among VNFs, ensuring interoperability and efficient interaction across the network. These APIs can pass metadata or tags indicating the criticality level of a service, ensuring that resource contention or failures are handled appropriately. By integrating criticality-aware metadata into API communications, the network can dynamically adapt to disruptions, reconfigure resources, and prioritize high-priority applications, ensuring their protection.
    
    By adopting a cloud-native approach, VNFs are decoupled from specific network sites, allowing for dynamic reallocation across distributed data centers. This flexibility mitigates the impact of localized failures, ensuring that critical services remain operational even under adverse conditions. For instance, \cite{7397878} focuses on the remediation and recovery phases of short-term resilience. In particular, the authors propose a proactive VNF failure restoration approach taking 3GPP specification into account for VNF relocation and lost state information restoration, as soon as a VNF starts malfunctioning.
    To truly assess the resilience of a cloud-native 6G network, the work \cite{wang2024cloud} proposes a Kubernetes-based framework that considers deep system failures as combinations of different instances and fault types.
    The work \cite{9089340} focuses on the availability of cloud-native service and management systems, e.g., the SMO in Fig.~\ref{fig:resilience_CPandManage_v2}.
    As a technique to enhance the resilience of cloud-based networks, \cite{oyeniran2024comprehensive} discusses service mesh techniques, which encounters features like load balancing, traffic management, automatic retries, and fault isolation. Under the premise of multiple cloud environments, the work \cite{9847048} explores the multi-cloud RAN overprovisioning and accounts for redundant CUs/DUs to serve a large-scale consumer network. From an O-RAN perspective, \cite{10329927} discusses the dynamic placement of RICs along the cloud-edge continuum, supported by an efficient strategy to react to sudden changes and dynamically re-deploy O-RAN components.

    To sum up, the integration of virtualization through VNFs and the adoption of cloudification with the O-Cloud architecture offer flexibility, adaptability, and significantly enhance network resilience through rapid reconfigurability and failure protection. Moving forward, the focus shifts to applying these principles in network slicing and QoS flows to effectively manage and prioritize mixed-critical traffic.

    \subsection{Network Slicing and QoS-Flows -- SDAP}\label{sec:SDAP}
    The 5G standard defines the SDAP protocol to map QoS flows from the 5G core to different data radio bearers, handling various types of traffic within logical pipelines based on their specific demands, such as QoS profiles \cite{dahlman20205g}. Moreover, network slicing divides a shared physical network into multiple virtual networks (slices), each tailored to specific service requirements. The key difference between these two concepts can be summarized as follows: Network slicing creates separate slices with specific configurations, while QoS flows within those slices manage traffic to meet service requirements.

    In the context of 6G, these concepts evolve significantly from their 5G roots to meet the more stringent demands of future networks \cite{9749222}. Beyond performance isolation and traffic prioritization, 6G's network slicing and QoS flows contribute directly to network resilience by ensuring priority and recovery mechanisms for critical traffic (i.e., \emph{mixed criticality}). They incorporate dynamic adaptability to changing network conditions (\emph{short- and long-term resilience}) and provide redundancy and failover mechanisms within slices and flows, supporting continuity even in the face of disruptions (\emph{robustness}).

    The use of distinct QoS flows enables flexible resource management tailored to individual flow requirements, which is crucial for deploying criticality-aware defense and adaptation mechanisms in the resilience framework. Under adverse conditions, such as network congestion or failures, QoS flows allow the network to prioritize critical applications. This guarantees that high-priority services, such as V2X, receive the necessary bandwidth and low latency, while non-critical flows can be deprioritized to protect safety-critical traffic \cite{9813735}.

    From a resilience perspective, network slicing allows each slice to have its own QoS parameters, resources, and policies. This creates a network-within-a-network, supporting different use cases, such as a slice for autonomous vehicles versus one for regular mobile users. Isolating critical services from non-critical ones within dedicated slices prevents disruptions in one slice from affecting the rest of the network. This isolation ensures that failures, such as overloads on one slice, do not impact other slices' services, particularly those hosting safety-critical applications.
    One such resilience approach is referred to as RAN slicing, where physical resource blocks are scheduled via the Near-RT RIC, as shown in Fig.~\ref{fig:resilience_CloudandVirtual}.
    Resources among different slices (inter-slice allocation) are managed through a common xApp, while intra-slice resource allocation (resources within each slice) is handled via separate xApps and the SDAP/PDCP stack \cite{9367527}. Inter-slice resource allocation can be static, ensuring strict slice isolation, or dynamic, allowing resource multiplexing to enhance efficiency \cite{Marquez2019ResourceSE}.
    
    Beyond robust slicing schemes, resilience and survivability are particularly enhanced by appropriate slicing configurations and prioritization schemes during massive outages (e.g., failure of multiple network nodes or links) \cite{10293082}.
    The framework in \cite{10293082} assigns applications to slices based on criticality and QoS (e.g., delay sensitivity vs. throughput sensitivity). Prioritization with different priority levels is shown to ensure a quick response to failures. Simulations evaluating \emph{absorption} and \emph{time-to-recovery} (see metric \eqref{eq:en}) highlight that the number of slices and the mapping of services are crucial to resilience. Interestingly, the specific slice configurations have a more significant impact than service prioritization schemes. However, no single slicing configuration performs optimally across all applications, necessitating flexible and dynamic slice management.\looseness-1

    Network slicing introduces a trade-off between resilience and resource costs. While it can protect critical services from failures in other slices (e.g., overloads), this isolation requires more resources as compared to managing services within a single slice using prioritization techniques. Under unpredictable, short-timescale traffic fluctuations, \cite{10107493} evaluates the effectiveness of isolation and auto-scaling techniques, such as network slicing and service prioritization in a 5G network. The results emphasize the importance of designing adaptation strategies that match traffic anomalies, ensuring that network service resilience is maintained.

    With multiple isolated virtual networks, each managing its own QoS flows to meet specific service levels, network slicing and QoS flows are key enablers for mixed criticality in 6G networks, ensuring resilience across diverse services. In the following subsection, we examine dual connectivity as well as hard- and software recovery mechanisms in the PDCP layer, which work in synergy with these resilience mechanisms to build a fully resilient 6G network.
    
    \subsection{Dual Connectivity and Failure Recovery -- PDCP}
    The PDCP layer is responsible for user data transmission, header compression, packet reordering, sequencing, and duplication. In 5G, the 3GPP defines a packet duplication feature at the PDCP layer based on dual connectivity \cite{8928163}.
    From both resilience and criticality perspectives, the PDCP layer plays a crucial role in real-time failure recovery, particularly for critical services. Firstly, through dual connectivity and packet duplication, the network can defend against and absorb anticipated failures on the radio link, mapping to the inner \emph{robustness} cycle of the proposed resilience strategy, see Fig.~\ref{fig:resilience_cycle}. Secondly, selective dual connectivity, which activates packet duplication only for critical services and network slices, enhances criticality-awareness within prospective 6G networks. Thirdly, adaptive dual connectivity and PDCP failure recovery mechanisms, supported by the reconfigurability and virtualization concepts discussed earlier, further enhance network resilience. For instance, by prioritizing critical data streams through dual connectivity during failures, the network can ensure low-latency recovery. These PDCP mechanisms strengthen the remediation and recovery phases (\emph{short-term resilience}) as well as the refine and fortification phases (\emph{long-term resilience}) as illustrated in Fig.~\ref{fig:resilience_cycle}.
    
    On an abstract level, dual connectivity enables a user to be connected to two distinct BSs operating on different carrier frequencies \cite{8928163}. Under this approach, two separate protocol sessions handle duplicate packet transmission, with the receiver processing the first packet to arrive and discarding the duplicate. However, in mixed-critical network environments, packet duplication is generally reserved for critical services and is activated, for instance, to support only URLLC services \cite{8928163}. The conventional dual connectivity can be viewed as a \emph{hot-backup} solution, where all packets are duplicated through fully redundant transmission. A more resource-efficient failure recovery mechanism, which also leverages dual connectivity, uses a backup fronthaul link that connects different virtual CUs and DUs \cite{9221182}. This framework, in addition to handling radio link failures, addresses hardware or software failures across the CU and DU. Compared to traditional fronthaul recovery methods, which exceed 5G and future 6G standards by requiring up to $10$s for recovery, this mechanism significantly reduces recovery times. By using PDCP filtering and selective packet duplication, CP and UP packets are duplicated; however, only the CP packets are sent from the virtual CUs to two distinct DUs, e.g., DU$1$ and DU$2$, as illustrated in Fig.~\ref{fig:resilience_CloudandVirtual}. This approach minimizes resource usage and allows for faster activation of the backup link in case of failure, reducing recovery times by up to $95$\% \cite{9221182}.

    In today's 5G dual connectivity mode, handling out-of-order packet arrivals on the PDCP layer presents a significant challenge. In such cases, the receiver must either wait for delayed packets to fill the sequence number gap or wait for the reordering timer to expire, which temporarily halts data delivery to the application \cite{9875285}. To address this limitation, the approach proposed in \cite{9875285} focuses on mitigating radio link failures and mobility events through the use of buffering, fast retransmission, and split activation mechanisms that dynamically adapt the PDCP layer at the transmitter. Results show that this solution can achieve near-zero interruption times with high reliability, thereby enhancing robustness and resilience.

    In conclusion, virtualization and cloudification enable flexible RAN deployment, support resilient operation, and ensure rapid error recovery across all protocol stack layers. This infrastructure also empowers network slicing, QoS flows, and adaptive, selective dual connectivity, which collectively enhance prioritization, criticality-awareness, and introduce various resilience mechanisms, which are crucial for resilient 6G services. However, the discussion thus far focuses primarily on the higher layers of the protocol stack. To achieve a holistic approach to criticality-aware and resilient 6G networks, it is essential to also consider the lower communication layers at the DU and RU, which we explore next.

    \section{Lower-layer DU and RU Schemes}\label{sec:lowlayer}
    While the upper layers of the protocol stack focus on maintaining application-level continuity and reliability, lower-layer resilience is centered on ensuring stable and consistent data delivery across the network infrastructure. This is achieved by incorporating redundancy, diversity, and dynamic adaptability. By leveraging path diversity, multi-connectivity, and automated error-handling mechanisms, these layers are equipped to rapidly respond to challenges such as congestion, packet loss, or hardware failures. Additionally, resilience is enhanced by integrating criticality-awareness into lower-layer protocols, allowing the network to prioritize essential services and ensure reliable performance even in the face of unforeseen disruptions.
    
    \begin{figure}[!t]
	\centering
	\includegraphics[width=\linewidth]{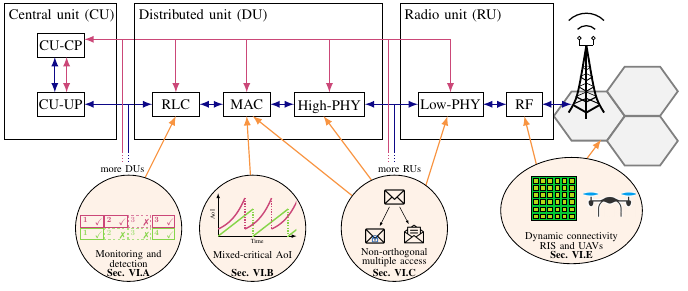}
	\caption{Resilience and criticality-aware concepts on the lower-layer 5G/O-RAN protocol stack, DU and RU.}
	\label{fig:resilience_DUandRU}
    \end{figure}
    Fig.~\ref{fig:resilience_DUandRU} shows the DU and RU components hosting the lower-layers of the protocol stack. More specifically, the CU-CP (or the O-RAN-defined RIC) controls and manages the operation of the UP lower-layers, which include the radio link control (RLC), medium access control (MAC), physical layer functions (PHY), and RF. In short, the RLC manages transmission reliability and flow control, the MAC controls access to physical communication channels, the PHY is responsible for transmitting bits over the physical medium (split into high and low functions, e.g., modulation and analog conversion), and the RF generates, amplifies, and transmits radio waves. Each of these layers can on its own contribute to the criticality-awareness and resilience prospects of the overall 6G network. In the following, this article reviews different technologies residing on the different layers, starting from the left hand side of Fig.~\ref{fig:resilience_DUandRU}. These technologies cover various aspects of resilient network design, i.e., error detection, prioritization, robust infrastructure, and adaptation.

    \subsection{Monitoring and Detection -- RLC}
    The RLC layer is located within the DU, below the PDCP layer and above the MAC layer \cite{etsiRLC}. Its main task is to segment service data units from the upper layers into \emph{smaller} protocol data units for further transmission and to reassemble them on the receiver side. There are three operational modes of the RLC, namely transparent, unacknowledged, and acknowledged mode. The transparent mode is used for non-reliable traffic, meaning no error correction mechanisms and no RLC processing, e.g., segmentation, are activated. In the unacknowledged mode, the RLC employs error correction and sequencing of data; however, the receiver does not acknowledge successful transmission. The acknowledged mode provides reliable data transmission by allowing acknowledgments and retransmissions. From a resilience and criticality perspective, RLC-based mechanisms such as, error correction and repeat requests, help sustain communication during dynamic or high-risk events, e.g., outages or interruptions in autonomous vehicles.

    In the context of resilience, referring to the inner cycle of Fig.~\ref{fig:resilience_cycle}, namely robustness, the RLC protocol layer contributes to sustaining short-term outages or sudden disruptions and helps anticipate erroneous transmissions to a certain extent. 
    This is achieved by supporting out-of-sequence packet receptions, allowing communication to continue despite temporary disruptions, with packet reordering handled at the PDCP layer \cite{9198680}. Additionally, forward error correction adds redundancy, such as repair bits, to transmitted packets \cite{roca2020rfc}, improving transmission reliability by detecting and autonomously correcting errors \cite{4741195}. However, out-of-sequence packet reception cannot fully mitigate complete packet loss, and error correction is effective only up to a certain packet loss rate.
    
    In mixed-criticality scenarios, the level of redundancy in forward error correction and the mode selection (between transparent and unacknowledged modes) must be tuned to meet varying service needs for reliability and latency. For instance, when ultra-low latency is critical, RLC can operate in transparent mode, introducing no overhead or delay. For higher reliability, the RLC should switch to unacknowledged or acknowledged mode \cite{7777932}. 6G use cases such as autonomous driving, where both latency and reliability are crucial, require a trade-off. In these cases, a cross-layer approach is necessary, where, for instance, the RLC operates in transparent mode, while other layers employ certain resilience mechanisms.

    For short-term resilience (second cycle in Fig.~\ref{fig:resilience_cycle}), the RLC is a central building block of the detection, remediation, and recovery phases. In the acknowledged operational mode, the RLC is able to detect transmission errors/outages. In more details, when a packet is lost, either the transmitter does not receive an acknowledgement, or the receiver sends a negative acknowledgement alongside an automatic repeat request. The transmitter would then retransmit the erroneous packets from the retransmission buffer \cite{9471818}. Within mixed-criticality scenarios, the trade-off between reliability and latency needs to be carefully managed. In an effort to adequately serve such prospective 6G applications, hybrid unacknowledged/acknowledged mode switching mechanisms have been proposed \cite{7777932}. That is, in transparent/unacknowledged mode, latency and processing power are reduced, while the overall throughput is increased; yet the communication is non-resilient. In contrast, the acknowledged mode enhances transmission robustness, detects errors, and remediates outages by using acknowledgments and retransmissions, but may increase the service latency.

    Overall, the RLC plays a crucial role in ensuring reliable data transmission in communication systems, particularly through mechanisms like out-of-sequence packet handling, forward error correction, and an automatic repeat request recovery. However, critical 6G applications require the careful management of RLC configuration in regards to the specific latency and reliability requirements, e.g., by hybrid mode switching techniques. Directly below the RLC layer, the MAC layer is responsible for scheduling, positioning it as a central component in managing criticality and enhancing resilience in dynamic 6G networks.

    \subsection{Criticality-Aware Scheduling -- MAC}    
    The MAC layer, as defined by 3GPP, maps logical channels to physical resources, intelligently managing these limited resources based on demand and priority levels to ensure efficient communication \cite{etsiMAC}. Prospective 6G applications, such as autonomous driving or XR, impose stringent requirements on both latency and reliability, requiring a specific focus on resilience and criticality management on the MAC layer. To address these needs, the MAC layer features a functionality called \emph{logical channel prioritization}, where priority parameters are provided by the RRC or Near-RT RIC. In terms of resilience, dynamic and adaptive scheduling policies allocate time and frequency resources in real time, allowing the MAC layer to respond to network fluctuations such as traffic load, user density, or channel degradation.

    One significant challenge associated with classical channel prioritization is the issue of \emph{starvation}. Starvation occurs when a lower-priority logical channel consistently receives insufficient access, often due to the persistent occupation of higher-priority channels. This phenomenon can result in substantial delays or even the complete halt of lower-priority service over a long time period. Additionally, current scheduling mechanisms in 5G networks are typically evaluated using metrics such as delay, throughput, goodput, fairness, spectral efficiency, and packet loss ratio \cite{9773317}. However, these state-of-the-art evaluation metrics are insufficient for capturing the critical demands of 6G, particularly when it comes to measuring and quantifying resilience, such as in the context of equation \eqref{eq:en}. 

    To integrate resilience and criticality awareness, MAC scheduling can incorporate the age of information (AoI) as a key metric. AoI measures the data freshness by capturing the time elapsed since the generation of a packet, e.g., a measurement in process control systems, until its successful delivery to the intended receiver. As resilience refers to a system's ability to withstand adverse conditions, absorb the impact of failures, restore the functionality, and recover from disruptions, a scheduling solution based on the AoI, such as the mixed-critical AoI in \cite{10279561}, holds great potential for enhancing network resilience. From an optimization perspective, the AoI considers the possibility of transmission failures while still functioning as a utility metric without becoming infeasible. Additionally, as a time-evolving metric, the AoI inherently seeks to minimize the data's age, which aligns with the integral functions in equation \eqref{eq:en}. For example, the work \cite{chaccour2020ruin} explores the occurrence of rare but extreme peak AoI values and proposes different scheduling policies to minimize such risk. The authors particularly show a trade-off between high average peak AoI values and minimized risk of ruin.
    While the AoI poses a promising utility metric for MAC scheduling in 6G, the choice of adaptation and recovery mechanisms remains critical to fully enable resilient services.

    In Fig.~\ref{fig:resilience_cycle}, short-term resilience consists of detect, remediate, and recover phases. To support autonomous failure detection, remediation of service functionality, and recovery, the 6G MAC layer can utilize a time-evolving metric, such as a mixed-critical AoI-based utility function \cite{10279561} or the data-rate-based approaches \cite{comebackkid}. 
    In mixed-critical networks, different weights can be assigned to each user's utility metric \cite{comebackkid}, or a linearly-evolving AoI can be used for non-critical users and exponentially-evolving AoI for safety-critical users \cite{10279561}. The latter distinction between linear and exponential AoI growth is illustrated in Fig.~\ref{fig:resilience_DUandRU}. Potential failures are detected automatically once the AoI becomes large, triggering an automatic prioritization mechanism, which allocates additional resources to critical users. Once the network recovers, the resource management can smoothly transition back to normal operations.

    In summary, by integrating time-evolving metrics such as AoI, the 6G MAC layer can dynamically detect, remediate, and recover from failures in mixed-critical networks. This ensures that critical services maintain their performance level through prioritized resource allocation during adverse conditions. The next section explores non-orthogonal multiple access (NOMA) both from a MAC and PHY layer perspective, for further enhancing resilience and criticality-awareness in 6G.

    \subsection{Non-Orthogonal Multiple Access -- MAC and PHY}\label{sssec:noma}
    While the MAC layer is primarily responsible for assigning time and spectrum resources to services, the PHY layer handles the transmission of bits over the physical medium through carefully chosen transmission and signaling methods. In complex and widespread future 6G networks, many users, devices, and services of mixed-critical nature (e.g., see Fig.~\ref{fig:hexagon}) share the same infrastructure. Together, both MAC and PHY layer determine how these multiple services can access the wireless communication medium, i.e., in terms of a multiple access scheme. Firstly, these schemes allow the limited resources to be robustly distributed among mixed-critical services, accounting beforehand to forseeable conditions such as network congestion, interference, or device density. Secondly, to mitigate the impact of unforseeable outages or disruptions, access techniques can be dynamically switched or reconfigured, and critical services can be given preferential access to resources.

    Traditional multiple access schemes are based on the orthogonal allocation of resources, separating users in frequency, time, and code domain \cite{dogra2022user}. In the context of network slicing (see Section \ref{sec:SDAP}), service guarantees for safety-critical or mission-critical applications are typically achieved through orthogonal network slicing. However, the allocation of orthogonal resource blocks often results in inefficient resource utilization due to a high risk of underutilization and overprovisioning of resources within a network slice. From a resilience point of view, such overprovisioned resources are then unavailable for other slices to utilize during resource scarcity or failure scenarios.

    In the face of 6G's increasing network density as well as the need for resilience and criticality-awareness, access schemes based on the non-orthogonal allocation of resources, e.g., NOMA, emerge as promising alternatives \cite{dai2018survey}. By allowing users to share the same time-frequency resources, NOMA and rate-splitting multiple access (RSMA), achieve substantially higher spectral efficiency, energy efficiency, fairness, robustness, scalability, and many more key performance indicators \cite{mao2022rate}.    
    RSMA's high potential for resilience is inherently present in the unique ability to partially decode interference and partially treat it as noise, which facilitates a flexible interference management approach, enabling dynamic adjustments to varying interference levels, channel conditions, and even unforseen errors \cite{comebackkid}. For criticality-awareness, (power-domain) NOMA leverages superposition coding, which allows multiple messages to be transmitted simultaneously by assigning different power levels. This enables the adjustment of each message's reliability, making NOMA particularly well-suited for mixed-critical networks \cite{karacora2024THzRIS, karacora2024intermittency}.

    Unlike orthogonal slicing methods, non-orthogonal schemes must address the challenge of dynamically allocating resources while still guaranteeing a certain performance level, particularly for high-criticality services. To tackle this, the concept of heterogeneous NOMA enables non-orthogonal resource sharing among users with different QoS demands \cite{8476595}. Leveraging the concept of \emph{reliability diversity}, which is closely related to mixed criticality, the authors in \cite{8476595} demonstrate that non-orthogonal multiplexing of heterogeneous services can indeed provide performance guarantees, and even outperform orthogonal slicing in certain regimes. More precisely, the inherent \emph{reliability diversity} of successive interference cancellation can be exploited by decoding critical data first and removing it from the received signal, before decoding the low-criticality data. We next discuss two concrete example how the above concept becomes a significant enabler for resilience in 6G.

    Consider a 6G network serving XR devices using THz technology. Due to the stringent XR demands for high data rates and real-time connectivity, data packets are classified based on their criticality. For instance, positional tracking and interactive feedback may be considered highly critical, while static XR scene elements might tolerate slight delays without compromising the overall user experience. In THz systems, faults are inevitable due to the trade-off between the intermittency of the dominant line-of-sight (LoS) path and the severe attenuation of alternative reflected paths \cite{karacora2024THzRIS, karacora2024intermittency}. By applying superposition coding to two mixed-critical data streams, more power is allocated to the critical data. Additionally, by utilizing the path diversity introduced by reflected paths, e.g., using RIS technology, critical data can be transmitted with higher reliability. Meanwhile, low-criticality data is efficiently transmitted whenever the strong LoS path is available. This approach, as demonstrated in \cite{karacora2024THzRIS}, is both robust and resource-efficient.

    Under the more general RSMA framework, the works in \cite{karacora2024resilience} and \cite{karacora2022ratesplitting} explore non-orthogonal uplink transmission of mixed-criticality data. Going beyond heterogeneous NOMA, RSMA splits each user's data stream, allowing for diverse reliability levels. This enhances RSMA's flexibility, supporting higher levels of \emph{absorption}, \emph{adaption}, and \emph{time-to-recovery}, as represented in the A$^3$RT-chart in Fig.~\ref{fig:resilience_1}. Moreover, since RSMA includes common message decoding across multiple receivers, it can enable multi-connectivity for high-criticality data mapped to the common stream, further boosting robustness. Taking one step further, RSMA's multi-connectivity capability strengthens short-term resilience by allowing more extensive resource reallocation for remediation and recovery processes \cite{comebackkid}. In terms of long-term resilience, refining common message decoding sets and device-connectivity can fortify overall network stability. While RSMA is applied to increase reliability with finite blocklength coding in \cite{karacora2022ratesplitting}, it also offers an efficient means of incorporating spatial macro-diversity to combat deep fading and channel blockages as shown in \cite{karacora2024resilience}.

    NOMA and RSMA emerge as flexible alternatives to traditional orthogonal schemes, designed to support resilience and criticality-awareness in 6G networks. NOMA enhances criticality management through \emph{reliability diversity}, while RSMA generalizes this approach, splitting data streams for diverse reliability levels and enabling multi-connectivity. This supports both short- and long-term resilience by facilitating dynamic resource allocation and exploiting spatial macro-diversity. Moving deeper into the protocol stack, we now explore how dynamic connectivity mechanisms integrate resilience and criticality at 6G's PHY and RF layers.
    
    \subsection{Dynamic Connectivity via RIS and UAVs -- PHY and RF}
    As a brief recap, the PHY layer handles the transmission and reception of raw data bits over the physical medium, managing aspects such as modulation, encoding, and synchronization. The RF layer processes data in the form of radio waves, handling tasks like frequency selection, signal amplification, filtering, noise reduction, and antenna management. To ensure a resilient wireless network infrastructure capable of withstanding node failures, deep fades, and blockages, both PHY and RF layers must provide robust connectivity through path diversity and redundancy. Additionally, criticality levels can be directly integrated at these lower layers. Techniques such as dense BS deployment, including C-RAN or distributed multiple-input multiple-output (MIMO) systems \cite{comebackkid}, as well as the integration of RIS \cite{10104574}, uncrewed aerial vehicles (UAVs) \cite{8757041}, and non-terrestrial networks (NTN) \cite{9080558} offer promising resilience enhancements. These technologies augment the existing infrastructure, enabling real-time adjustments and optimizations in response to dynamic environmental conditions and user requirements.

    The C-RAN architecture relies on a centralized processor with baseband processing capabilities, coupled with multiple distributed BSs of lower complexity \cite{peng2016recent}. At the PHY layer, distributed or cell-free MIMO allows these BSs to jointly serve users via coherent signal processing \cite{bjornson2020scalable}. Referring to the resilience cycle in Fig.~\ref{fig:resilience_cycle}, the dense deployment of BSs within C-RAN networks inherently provides multi-connectivity, making the network robust against local BS outages and poor channel conditions. This increased connectivity ensures that remediation efforts have access to a larger pool of resources to counteract failure impacts. Centralized processing is a key enabler for this architecture, providing the ability to anticipate, detect, and diagnose failures while analyzing previous resilience cycles. Additionally, centralized management assesses global network resources for remediation and recovery, while utilizing global network knowledge to refine and strengthen long-term operations. For example, \cite{comebackkid} highlights the effectiveness of resilience strategies in mixed-critical C-RAN networks, showing that user-to-BS association during the remediation phase can significantly enhance recovery, even under severe faults.

    While C-RAN represents a comprehensive network architecture designed to enhance resilience through centralized processing and distributed base stations, RIS operates more as a flexible enhancement that can be integrated into various architectures, including C-RAN. By augmenting existing infrastructures, RIS provides additional layers of adaptability and robustness, particularly in scenarios where traditional network paths face blockages or impairments. RIS consist of metasurfaces made up of numerous tunable reflecting elements that can introduce real-time phase shifts to incoming signals. The adaptability of a RIS allows it to extend the network coverage, suppress interference, and alter channel statistics \cite{liu2021reconfigurable}. From a resilience standpoint, RIS offer dual advantages. First, they introduce new \emph{redundant} paths to the system, which can be utilized in the event of blockages in direct paths. Second, these new paths are \emph{dynamically} customizable, enabling rapid adaptation to disruptions through intelligent phase-shifter configurations.
    Due to channel impairments and blockages, RIS technology serves as a backup strategy at the PHY/RF layer, enhancing the networks resilience \cite{10104574,sivadevuni2023preparing}.
    The work in \cite{10104574} proposes an alternating optimization problem, where beamforming vectors, rate allocations at access points, and RIS phase shifts are jointly designed to improve resilience, as defined by equation \eqref{eq:en}. However, for time-critical services, the complexity of optimizing RIS phase shift configurations might be infeasible within the  required time constraints. Interestingly, the results in \cite{10104574} demonstrate that even an unoptimized (randomly initialized) RIS already increases the network resilience utility. The work \cite{sivadevuni2023preparing} demonstrates the use of joint communication and sensing in a RIS-aided network. By sensing line-of-sight link blockages in advance, it enables a proactive RIS-based recovery technique. Similarly, when accounting for criticality, an alternative RIS path serves as a weaker, yet more stable, backup path in case the primary LoS path is blocked. In such scenarios, critical data can be transmitted over both the RIS and LoS paths, while low-criticality data is transmitted only when the LoS path is available, ensuring efficient resource use and enhanced reliability for critical services \cite{karacora2024THzRIS}.

    While RIS provide passive, adaptable path enhancements through intelligent signal reflection, UAVs offer active, mobile network elements that can be dynamically deployed to enhance coverage, reliability, and resilience in real-time. Unlike RIS, UAVs can physically reposition themselves to address connectivity challenges, making them especially valuable in rapidly changing or hard-to-reach environments. In general, a UAV is an aircraft that operates without a human pilot on board, controlled either remotely or autonomously. From a resilience perspective, UAVs provide redundant communication paths, can carry redundant lightweight BSs, extend the network diversity via an aerial link, and can dynamically change their position and configuration. While the work \cite{8757041} utilizes the fast adaptability of UAVs, \cite{10468637} considers a mixed-criticality setup and utilizes the UAV trajectory design.
    In more details, in \cite{8757041}, a UAV-assisted network under BS breakdown scenarios is considered. When a ground BS fails to operate properly, this framework deploys a backup UAV to take over parts of the BS's network operations. Specifically, the RSMA scenario in \cite{8757041} is as follows: the remaining active BSs transmit private messages to their connected users, while the UAVs transmit common messages only. On one hand, the users who become underserved due to the BS breakdown can be reconnected using the UAVs. On the other hand, the UAV-induced heavy interference is mitigated using the RSMA common message decoding scheme. The results in \cite{8757041} show that the recovered sum-rate is significantly increased compared to the no-RSMA and no-UAV schemes. Furthermore, the work in \cite{10468637} takes this a step further by accounting for UAV mobility in the context of resilient disaster relief through UAV trajectory design. With a focus on mixed criticality, the proposed framework is specifically able to distinguish between reduced capacity and mission-critical users. The results in \cite{10468637} demonstrate how the proposed methods achieve increased satisfactory connections in disaster relief scenarios.
    Taking a step beyond UAVs and into the stratosphere, high-altitude platforms and satellite stations provide global connectivity and help fill underserved coverage gaps \cite{9080558}. These NTNs can also enhance resilience; for example, \cite{9089385} demonstrates how NTNs can reroute failed user demands, significantly improving overall network survivability. Despite increasing attention from the research community, further research is needed to optimize NTNs' integration into terrestrial networks, enhance their resilience capabilities, address latency challenges, and enable criticality-aware resource management.

    In essence, the lower layers of the protocol stack, as illustrated in Fig.~\ref{fig:resilience_DUandRU}, form the foundation for ensuring reliable and stable data delivery in a 6G network, incorporating resilience and criticality-awareness. These layers enable the network to adapt to adverse conditions and prioritize essential services. For instance, critical services can employ the RLC acknowledged transmission mode, while mixed-critical data streams may be managed through AoI-aware scheduling algorithms to maintain data freshness and system responsiveness. In terms of access techniques, NOMA and RSMA can inherently accommodate mixed-criticality by prioritizing critical data during decoding. Moreover, dynamic and reconfigurable solutions like RIS, UAVs, and NTNs can effectively address path failures, deep fades, and blockages, further enhancing the network's ability to recover and maintain service continuity.
   
 \section{Open Research Directions}\label{sec:outlook}
    \subsection{Resilience Metrics}
    As a fundamental goal, 6G networks aim to integrate resilience across all communication layers and components. However, the lack of precise and universally accepted resilience metrics remains a significant barrier. While traditional metrics, such as packet delivery ratio or latency, provide a baseline, they fail to comprehensively capture the multi-faceted nature of resilience, which spans reliability, robustness, adaptability, and recovery across diverse scenarios. Developing comprehensive resilience metrics that account for various factors and trade-offs, such as performance, cost, and service differentiation, is crucial, as accurate evaluation of resilience forms the foundation for designing effective policies. The lack of standardized metrics also complicates comparisons between proposed solutions.
    Future research must focus on developing resilience metrics that encapsulate both qualitative and quantitative dimensions. For instance, metrics could include recovery time from failures, adaptive throughput under dynamic conditions, or criticality-weighted service continuity to reflect the differentiated priorities of mixed-criticality systems. In \cite{10716652}, some related discussions capture the mean-time-to-detect, -remediate, -recover, and -adapt.

    Moreover, these metrics should align with the phases of resilience, i.e., the A$^3$RT-chart in Fig.~\ref{fig:resilience_1} and equation \eqref{eq:en}, while considering trade-offs, such as energy efficiency and latency. Incorporating AI/ML-driven methods to dynamically compute and update these metrics in real time could also prove transformative, enabling networks to self-assess their resilience continuously. By establishing precise, scenario-specific, and scalable metrics, the research community can better quantify progress, unify evaluation standards, and guide the development of truly resilient and criticality-aware 6G systems.
 
    \subsection{Artificial Intelligence and Decentralized Learning}
    Based on the above discussions, AI/ML-based solutions already play an important role in the envisioned 6G network protocol stack, specifically within the O-Cloud SMO block. These solutions are crucial for network analysis, prediction, and the implementation of adaptable decision policies. Thanks to their generalization, robustness, and learning capabilities, AI technologies serve as major enablers for enhancing resilience, covering all phases of the resilience cycle illustrated in Fig.~\ref{fig:resilience_cycle}. However, it is important to note that current AI tools struggle to cope with unforeseen scenarios, posing a significant challenge for resilient network operation. Many narrow AI systems are designed for a specific set of tasks and lack the flexibility to adapt to new conditions or perform tasks outside their predefined scope. In fact, one can state that existing AI systems are themselves not ``resilient'' to unseen or unfamiliar data points, as they require overhead expensive re-training or massive datasets for offline training. The recent work \cite{saad2024artificial} presents a vision for artificial general intelligence (AGI)-native 6G systems, that can address those shortcomings by leveraging common sense and incorporating components for perception, world modeling, and action planning. AGI systems, unlike narrow AI, are capable of understanding, learning, and mirroring human intelligence across various contexts \cite{10599272}. For example, AGI's \emph{reasoning} capabilities allow it to infer cause-effect relationships within network data and extend these to different logical conclusions, reducing the need for constant re-training \cite{thomas2023reliable}. Additionally, AGI's \emph{planning} ability enables it to devise intermediate steps toward greater network objectives, such as ``reserving resources while ensuring that each link operates below $50$ \% capacity," \cite{9925251}. These AGI-native systems have the capability to adequately generalize and abstract real-world concepts, ultimately facilitating the development of a truly resilient 6G network.
 
    Another key challenge for AI/ML-based resilience solutions is that the data required for ML training is becoming increasingly distributed across multiple agents (e.g., 5G protocol units such as CU, DU), due to constraints like memory and computing availability, and privacy \cite{9084329}. To address this, decentralized learning emerges as a candidate solution with wide-ranging resilience facets such as alleviating single-points-of-failure and reducing the computational requirements at single nodes. Additionally, 6G use cases like autonomous driving, industrial automation, and healthcare particularly benefit from decentralized approaches, as they preserve privacy, ensure scalability, and enable low-latency, local decision-making. Multi-agent reinforcement learning in 6G enables intelligent agents to make decisions by interacting with dynamic environments, including infrastructure and devices. Through trial and error, these agents enhance network adaptability and efficiency \cite{zhu2024survey}. A key challenge is the rapidly changing nature of 6G, particularly in areas like autonomous driving, where fast decision-making is crucial for resilience. Similarly, federated learning allows devices to collaboratively train a shared model while keeping data local, though it faces issues like high communication costs and slow convergence \cite{9205981}. Therefore, AGI and decentralized learning become increasingly important subjects of study for resilience within the context of 6G.

    \subsection{Resilient Near-Field Beamfocusing}
    In 6G, operating in mmWave and THz bands requires larger antenna arrays to combat spreading losses and ensure performance over wider bands while minimizing transmit power. The increased antenna aperture and higher frequencies expand the near-field zone around the transmitter, where beamfocusing offers greater flexibility in system design. This makes near-field beamfocusing critical for optimizing 6G network performance \cite{10438970}. One novel approach at the PHY and RF layers is wavefront hopping, similar to frequency hopping, allowing transmitters to switch between preconfigured wavefronts to maintain resilience against blockages \cite{10438970}.
    
    Recent advancements in RIS-based wavefront engineering have shown that RIS-enabled beam profiles are naturally resilient, offering the ability to focus, bend, and even self-heal \cite{stratidakis2023perceptive}. These innovations highlight RIS's potential for creating perceptive, resilient, and efficient network solutions. Additionally, hybrid beamforming/beamfocusing schemes are being explored for improved localization techniques. Despite the promising developments, resilient near-field beamfocusing is still in its early stages, signaling a significant research opportunity in future 6G networks.

    \subsection{Semantic and Task-Oriented Approaches}
    As discussed previously, maintaining awareness of the criticality of services is essential for enabling resilient communication networks. Taking this a step further, the advantages of accounting for the significance and semantic meaning of individual data packets within the realm of resilient communication is worth investigating. Semantic and task-oriented communication has emerged as a promising approach for 6G and beyond communication networks due to its expected efficiency and ability to enhance user experience.  
    Building on recent proposals, the concept of a semantic-aware O-RAN architecture offers promising directions for enhancing both user and control plane functions within 6G \cite{li2023open, strinati2024goal}. A dedicated `semantic plane' could be introduced to refine communication by focusing on task-specific feature extraction and relevance-based encoding. This shift towards semantic communication presents opportunities for research in efficient data compression and in minimizing control plane overhead, such as by prioritizing features of channel state information most relevant to network performance \cite{li2023open}.
    Future work could explore the implementation of a semantic RIC within the O-RAN framework, orchestrating semantic applications to optimize network functions dynamically. Additionally, a `semantic engine' integrated within the SMO layer could manage semantic processing resources and oversee the development lifecycle of these models. By advancing these ideas, semantic-aware architectures could significantly improve network resilience, adapting source coding to channel conditions and supporting information recovery even with partial data loss, thus enabling reliable, critical information delivery amid network disruptions \cite{li2023open}.
    However, deploying semantic communication strategies requires a different view on resilience, as various encoded feature units may have different impacts on the semantic inference at the receiver. Hence, particularly sensitive features should be prioritized, e.g., by assigning them to high-quality orthogonal frequency-division multiplexing subchannels as proposed in \cite{channel_resilience}. 
    In addition to AI-based semantic communication, strategies that tailor transmission methods according to data significance show great promise. These approaches not only reduce unnecessary traffic but also enhance the robustness of critical or vulnerable data transmissions \cite{karacora2024THzRIS}.
 
    In essence, rather than focusing solely on stable connectivity and bit delivery, resilience should be addressed from a semantic perspective, prioritizing task completion over mere data transmission. Such a goal-oriented approach is particularly promising for ensuring resilience and reliable system functionality in mission- and safety-critical applications, e.g., autonomous driving, telemedicine, or industrial automation, where the accurate and timely delivery of data is paramount to the operation's success and safety.

    \subsection{Trustworthiness and Integrity}
    Ensuring security and trustworthiness is a significant challenge for resilient 6G communication networks. The innovative features of next-generation wireless networks, such as virtualization, cloud computing, AI/ML, and increased dynamic adaptability, introduce complex security challenges. While these advancements offer substantial opportunities for network resilience, as discussed in previous sections, they also create new avenues for security threats, targeting virtualization functions, network configuration modules, and AI algorithms for QoS management \cite{sedjelmaci2024secure}.
    
    While the open and disaggregated architecture of O-RAN enables the flexibility needed for resilient network operation, the softwarization of network functions introduces security vulnerabilities. Specifically, VNFs rely on shared third-party cloud infrastructure for information exchange across protocol layers, which increases the risk of exploitation, e.g., attackers taking control of network nodes \cite{ramezanpour2022intelligent}. As a result, adopting a \emph{zero-trust} principle emerges as a vital solution for resilient 6G networks. By treating trust as a potential vulnerability, this approach involves assuming no inherent trust even within the network perimeter, necessitating continuous verification and stringent access controls. Furthermore, given the inherent adaptability of future communication protocols and infrastructure, which is crucial for resilient connectivity and effective failure response, traditional static security measures may become insufficient, requiring adaptive, real-time security mechanisms.
    
    The promising advantages of learning-based network management in an AI-native communication system, e.g., enabling self-healing networks and digital twins, also present challenges, such as ensuring data integrity and protecting AI models. Additionally, the cloud-native architecture with centralized RAN orchestration creates central points of failure and targets for adversarial attacks. Therefore, specific strategies are required to protect the network functionality, including secure data storage and confidential computing in the cloud \cite{nextGATrust}.
  
    \subsection{Scalability and Sustainability}
    Given the massive connectivity expected in 6G networks, resilience policies and error response mechanisms must scale to accommodate large numbers of devices while maintaining consistent performance and avoiding congestion \cite{kalor2024wireless}. A key aspect of resilient 6G is adaptive and dynamic resource management, where intelligent algorithms process vast datasets in real-time. Developing low-complexity, computationally efficient solutions is essential to enhance both scalability and energy efficiency while ensuring swift responses to changing network conditions.

    In light of the global climate crisis, sustainability becomes critical \cite{nextGAroadmap}. Reducing the carbon footprint of 6G systems demands careful resource management alongside resilient design. Balancing resource-intensive measures, like redundancy and diversity, with demand-driven response mechanisms is crucial for building a resilience framework that remains efficient and energy-conscious. The work in \cite{10694673} offers an initial step, proposing an energy-efficient private message removal mechanism for congested RSMA-based networks.

    By incorporating scalability, sustainability, and resilience, 6G networks can maintain high reliability and performance while minimizing energy consumption and managing the complexity of vast device ecosystems.
 
 \section{Conclusion}\label{sec:con}
 This article integrates the concepts of resilience and criticality within the context of future 6G communication networks, reviewing their individual definitions, discussing their applications in existing communication networks, and presenting their prospective synergies. In particular, 6G poses unique challenges such as unforeseen errors, mixed-critical service requirements (including safety-criticality), decentralization, virtualization, and interdependent network components. Addressing these challenges requires comprehensive resilience across all protocol layers, encompassing error anticipation and detection, defense, adaptation, and remediation strategies, as well as continuous performance analysis. 
 On the control and management plane, AI-native implementations of RIC components and the integration of resilient and criticality-aware resource management schemes in the Near-RT RIC and RRC become major enablers for resilient 6G networks. Virtualization and cloudification of the O-Cloud, network slicing, and recovery mechanisms at higher protocol layers significantly contribute to resilience. At the lower communication layers, acknowledged transmission, criticality-aware scheduling, NOMA and RSMA-based access schemes, and a controllable physical layer (e.g., via RIS and UAVs) promise to enhance 6G resilience. Looking forward, future works may focus on AGI, semantic approaches, trustworthiness, sustainability, as well as comprehensive inter-layer schemes. 
 To conclude, four key lessons learned for 6G resilience and mixed criticality are: $(a)$ Resilience strategies must be integrated across all layers of the 6G protocol stack, from the PHY at the air interface to the management and control plane and the core network; $(2)$ Differentiating between critical and non-critical services is essential for efficient resource management and prioritization; $(3)$ Resilience is dynamic, the ability to adapt and reconfigure network resources, including through AI-native devices, is vital for maintaining service continuity during unforeseen conditions; $(4)$ The adoption of VNFs, modular architectures, cloud-native frameworks, criticality-aware scheduling, adaptable access schemes, and dynamic PHY techniques provides the necessary flexibility to efficiently handle failures and disruptions.
 To effectively support mixed-critical applications despite unforeseeable failures and resource scarcity, all these efforts hinge on seamlessly integrating resilience and criticality as brothers in arms for 6G.


\bibliographystyle{IEEEtran}
\bibliography{bibliography/references}



\begin{IEEEbiographynophoto}{Robert-Jeron Reifert} (S'19) received the B.Sc. and M.Sc. degree in Electrical Engineering and Information Technology from Ruhr University Bochum, Germany, in 2019 and 2021, respectively. He is one of the recipients of the Association for Electrical, Electronic and Information Technologies (VDE) Rhein-Ruhr graduate student award 2021. He is currently pursuing the Ph.D. degree with the Institute of Digital Communication Systems, Ruhr University Bochum, Germany. His research interests include wireless communication systems, mixed criticality, and resilience in 6G communication networks and beyond.
\end{IEEEbiographynophoto}
\begin{IEEEbiographynophoto}{Yasemin Karacora} (S'17) received her B.Sc. and M.Sc. degree in Electrical Engineering and Information Technology from Ruhr University Bochum, Germany, in 2016 and 2019, respectively. From 2016 to 2017 she was an exchange student at the ECE Department of Purdue University, IN, USA. She is currently pursuing her Ph.D. degree at the Institute of Digital Communication Systems, Ruhr University Bochum, Germany. Her research interests include wireless communication, beamforming at (sub-)terahertz frequency bands, and reliability and resilience in 5G and 6G networks.
\end{IEEEbiographynophoto}
\begin{IEEEbiographynophoto}{Christina Chaccour} (S'17-M'23) is currently a Network Solutions Manager at Ericsson, Inc. In her role, Dr. Chaccour seamlessly bridges strategy, product solutions, and research, spearheading developments in 5G Advanced, 6G networks, and AI-integrated solutions while ensuring the responsible and innovative use of AI, addressing both policy and technical dimensions. Dr. Chaccour actively represents Ericsson as a delegate in prominent industry bodies and councils including FCC CSRIC IX, 5G Americas, and Next G Alliance. With a Ph.D. in Electrical Engineering from Virginia Tech, her academic journey yielded significant contributions, particularly in 6G systems at THz frequencies for next-gen XR and holographic systems, her work also laid the groundwork for pioneering research in AI-native networks and semantic communications. Her dissertation contributions and their subsequent impact were recognized with the Bill and LaRue Blackwell Award.
Dr. Chaccour has been recognized with several honors, including the Best Paper Award at the 10th IFIP Conference on New Technologies, Mobility, and Security (NTMS) in 2019 and the Exemplary Reviewer Award from IEEE Transactions on Communications in 2021, a distinction given to fewer than 2\% of reviewers. Her paper in IEEE Communication Surveys and Tutorials was featured in the Top Access article listing from June to November 2022. In 2024, she was named among the ``Top 100 Brilliant and Inspiring Women in 6G."
Dr. Chaccour serves on the editorial board of IEEE Transactions on Machine Learning in Communications and Networking, IEEE Transactions on Cognitive Communications and Networking, Wireless Personal Communications (Springer), and the guest editorial board of IEEE Communications Standards Magazine within the series ``AI for Wireless." She is also on the advisory board of ``The Data Science Conference".
In her entrepreneurial endeavors, Dr. Chaccour co-founded the startup ``Internet of Trees," which has garnered numerous local and international awards.
\end{IEEEbiographynophoto}
\begin{IEEEbiographynophoto}{Aydin Sezgin} (S'01-M'05--SM'13) received the Dr. Ing. (Ph.D.) degree in electrical engineering from TU Berlin, in 2005. From 2001 to 2006, he was with the Heinrich-Hertz-Institut, Berlin. From 2006 to 2009, he held postdoctoral positions with the Information Systems Laboratory, Stanford University, USA and the Department of EECS, University of California, Irvine, USA. He is currently a professor with the Ruhr University Bochum, Germany. He has published several book chapters more than 65 journals and 200 conference papers in these topics. Aydin is a winner of the ITG-Sponsorship Award, in 2006. He was a first recipient of the prestigious Emmy-Noether Grant by the German Research Foundation in communication engineering, in 2009. He has coauthored papers that received the Best Poster Award at the IEEE Communication Theory Workshop, in 2011, the Best Paper Award at ICCSPA, in 2015, and the Best Paper Award at ICC, in 2019. 
\end{IEEEbiographynophoto}
\begin{IEEEbiographynophoto}{Walid Saad} (S'07, M'10, SM'15, F'19) received his Ph.D degree from the University of Oslo in 2010. He is a Professor at the Department of Electrical and Computer Engineering at Virginia Tech where he leads the Network sciEnce, Wireless, and Security (NEWS) laboratory. His  research interests include wireless networks, machine learning, game theory, cybersecurity, drones, semantic communications, and cyber-physical systems. Dr. Saad was the author/co-author of eleven conference best paper awards, of the 2015 and 2022 IEEE ComSoc Fred W. Ellersick Prize, and of the 2023 IEEE Marconi Prize Paper Award. He is a Fellow of the IEEE.
\end{IEEEbiographynophoto}

\end{document}